\documentclass[aip,amsmath,amssymb,reprint,a4paper,superscriptaddress]{revtex4-1}

\usepackage{xspace}
\usepackage{rotating}
\usepackage{dcolumn}
\usepackage{xr}
\usepackage{siunitx}
\usepackage{pstool}
\usepackage{booktabs}
\usepackage{setspace}
\usepackage{tabulary}
\usepackage{array}
\usepackage{overpic}

\newcommand{\kin}{\mathbf{k}_\mathrm{in}}
\newcommand{\kout}{\mathbf{k}_\mathrm{out}}

\let\vec\mathbf   

\DeclareSIUnit\angstrom{\text {Å}}

\newcommand{\cfeldesy}{\affiliation{Center for Free-Electron Laser Science CFEL, Deutsches
    Elektronen-Synchrotron DESY, Notkestr.\ 85, 22607 Hamburg, Germany}}
\newcommand{\uhhcui}{\affiliation{The Hamburg Center for Ultrafast Imaging, 
    Luruper Chaussee 149, 22761 Hamburg, Germany}} 
\newcommand{\uhhphys}{\affiliation{Department of Physics, University of Hamburg, Luruper
    Chaussee 149, 22761 Hamburg, Germany}}

\begin{document}


\title{Convergent-beam attosecond X-ray crystallography} 

\author{Henry N. Chapman}\email[Correspondence to: ]{henry.chapman@desy.de.}\cfeldesy\uhhcui\uhhphys
\author{Chufeng Li}\cfeldesy
\author{Sa\v{s}a Bajt}\email[]{sasa.bajt@desy.de}\cfeldesy\uhhcui
\author{Mansi Butola}\uhhcui
\author{J. Lukas Dresselhaus}\uhhcui
\author{Dmitry Egorov}\cfeldesy
\author{Holger Fleckenstein}\cfeldesy
\author{Nikolay Ivanov}\uhhcui
\author{Antonia Kiene}\cfeldesy
\author{Bjarne Klopprogge}\uhhcui
\author{Viviane Kremling}\cfeldesy
\author{Philipp Middendorf}\cfeldesy
\author{Dominik Oberthuer}\cfeldesy
\author{Mauro Prasciolu}\cfeldesy
\author{T. Emilie S. Scheer}\cfeldesy
\author{Janina Sprenger}\cfeldesy
\author{Jia Chyi Wong}\uhhcui
\author{Oleksandr Yefanov}\cfeldesy
\author{Margarita Zakharova}\cfeldesy
\author{Wenhui Zhang}\cfeldesy

\date{\today}

\begin{abstract}
Sub-ångstr\"{o}m spatial resolution of electron density coupled with sub-femtosecond
temporal resolution is required to directly observe the dynamics of the electronic
structure of a molecule after photoinitiation or some other ultrafast
perturbation. Meeting this challenge, pushing the field of quantum crystallography to
attosecond timescales, would bring insights into how the electronic and nuclear degrees of
freedom couple, enable the study of quantum coherences involved in molecular dynamics, and
ultimately enable these dynamics to be controlled. Here we propose to reach this realm by
employing convergent-beam X-ray crystallography with high-power attosecond pulses from a
hard-X-ray free-electron laser. We show that with dispersive optics, such as multilayer
Laue lenses of high numerical aperture, it becomes possible to encode time into the
resulting diffraction pattern with deep sub-femtosecond precision. Each snapshot
diffraction pattern consists of Bragg streaks that can be mapped back to arrival times and
positions of X-rays on the face of a crystal. This can span tens of femtoseconds, and can
be finely sampled as we demonstrate experimentally.  The approach brings several other
advantages, such as an increase of the number of observable reflections in a snapshot
diffraction pattern, all fully integrated, to improve the speed and accuracy of serial
crystallography---especially for crystals of small molecules.
\end{abstract}


\maketitle 

\section{Introduction}
\label{sec:intro}
Time-resolved serial femtosecond crystallography has opened up the domain of ultrafast
structural biology and chemical dynamics~\cite{Branden:2021,Khusainov:2024}. By utilising
intense femtosecond-duration pulses from X-ray free-electron lasers, complete sets of
structure factors of crystals of photoactive macromolecules have been recorded at a series
of delay times after reaction initiation using an ultrafast optical laser pulse. In this
way, for example, the nuclear motions in processes such as photosynthesis in a
photosynthetic reaction center~\cite{Dods:2021}, light sensing in photoactive yellow
protein~\cite{Pande:2016} and
rhodopsins\cite{Nogly:2018,Gruhl:2023}, the response of a photoswitchable fluorescent
protein~\cite{Coquelle:2018}, and DNA repair by
photolyase~\cite{Maestre-Reyna:2023,Christou:2023} have been observed on femtosecond
timescales. The method has also recently been applied to measure the response of a
light-sensitive metal-organic framework structure~\cite{Kang:2024}.  Although the number
of known photoactive proteins and other molecules is comparatively small, these studies
may offer direct insights into chemistry and allow the validation of quantum chemistry
codes. The nuclei motions, however, act in response to excited-state electron dynamics
initiated by the absorption of a photon. Having much lower mass, the dynamics of electrons
occur on even faster timescales. So far, spectroscopic methods have been the only route to
investigate these rapid changes~\cite{Calegari:2014}. A real-space visualisation of
structural dynamics on sub-femtosecond to few-femtosecond time scales is required to
improve our understandinqg of the evolution of the excited system and the coupling of
electronic and nuclear motions in molecules. For example, it could enable the observation of the
non-adiabatic evolution of dynamics of an excited molecule, where the superpositions of wave packets on
the potential-energy surfaces may lead to interferences that may route the electrons and
nuclei back to the group state or to various product states.
With the recent development of
sub-femtosecond duration X-ray pulses with terawatt powers~\cite{Franz:2024}, it becomes
feasible to conduct crystallography in the attosecond regime. Such an undertaking comes
with several challenges. High-precision crystallographic measurements are required at
resolutions better than the atomic scale of \SI{0.7}{\angstrom} or so to be able to
characterise weak changes of electron density, as is achieved in the field of quantum
crystallography~\cite{Genoni:2018} and recently demonstrated in static serial
crystallography experiments on rhodamine~\cite{Takaba:2023} and in femtosecond X-ray
powder diffraction of charge relocations in terahertz excitation of aspirin crystals~\cite{Hauf:2019}.
Reaction initiation with sub-femtosecond precision necessarily demands wavelengths in the
deep UV or shorter, used to generate attosecond pulses. Path differences between pump and
probe beams must be controlled to below about \SI{30}{\nano\meter}. Addressing these
challenges would no doubt help further develop and improve the study of time-resolved
structures  also at longer timescales.

Here, we propose and explore a route based on convergent-beam diffraction, utilising a
highly-focused beam, to address some of these challenges. The approach could greatly
increase the information content of single snapshot patterns by providing many more Bragg
reflections than with a collimated monochromatic beam, and ensuring that most of those are
fully integrated, thereby enabling the acquisition of datasets with higher precision or
with fewer diffraction patterns.  Since the method combines aspects of projection imaging,
effects of crystal shape and structure can be inferred and accounted for, improving the
accuracy of the dataset. A particularly useful characteristic of convergent-beam
diffraction that we examine here is that time can be directly encoded in the angular
distribution of diffracted intensities, analogous to the way that chirped-pulse Laue
diffraction maps time to the wavelength of reflections as proposed by Keith
Moffat~\cite{Moffat:2003}. Furthermore, with high-quality focusing optics, the necessary
wavefront control for sub-femtosecond timing should be achievable.  Following a brief
overview of serial crystallography and its extension to pink-beam measurements in
Sec.~\ref{sec:serial}, convergent-beam diffraction of three-dimensional crystals is
introduced in Sec.~\ref{sec:CB-diff} and compared with Laue diffraction. Two particular
schemes are considered: one in Sec.~\ref{sec:in-focus} where the crystal is located in the
focal plane of the X-ray lens, and the other in Sec.~\ref{sec:out-of-focus} where it is
placed out of focus. In both cases, time can be encoded directly in the diffraction
patterns when using a dispersive focusing lens, as explained in Sec.~\ref{sec:TR}. The
latter scheme enables highly magnified topograms of the crystal to be obtained, from which
the X-ray arrival time can be determined. For visible or UV pump pulses, we must consider
that the refractive index of the crystal material is considerably different than for
X-rays, leading to different speeds of propagation through the crystal. Particular
velocity-matching schemes are required, as described in Sec.~\ref{sec:matching}. These
constrain the orientations of the crystal and pump pulse relative to the X-ray beam that
can be used to perform time-resolved crossed-beam topography~\cite{Neutze:1997}, but
nevertheless allow diffraction data to be obtained over a span of tens of femtoseconds in
a single snapshot pattern while maintaining sub-femtosecond time resolution. We present an
experimental test of convergent-beam diffraction on a vitamin B$_{12}$ crystal in
Sec.~\ref{sec:B12} using an X-ray lens of very high numerical aperture. In
Sec.~\ref{sec:resolution}, we show that our particular experimental geometry would encode
time in the diffraction pattern to a resolution approaching \SI{10}{\atto\second} over a
range of \SI{12}{\femto\second}, and describe how this can be adjusted.

\section{Serial Femtosecond Crystallography}
\label{sec:serial}
Serial femtosecond crystallography is a method where many snapshot (or still) diffraction
patterns are collected, each from an individual crystal that is usually in some random and
unknown orientation~\cite{Barends:2022}. The patterns are indexed and intensities in Bragg
peaks are extracted and merged, ideally after first accounting for their
partialities~\cite{White:2016}. The method can be compared with powder diffraction, where
intensities from many individual crystallites are integrated in the laboratory frame over
Debye-Scherrer rings. By recording diffraction instead from each single crystallite, the
amalgamation of diffraction intensities from many crystals can be performed in
three-dimensional reciprocal space, in the frame of reference of the crystal lattice
rather than the laboratory frame, to obtain the complete set of structure factors of the
average three-dimensional crystal. By using femtosecond-duration high-intensity X-ray
free-electron laser pulses, exposures may greatly exceed the usual limits set by radiation
damage of macromolecular crystals, even at ambient temperatures, by outrunning damage
processes~\cite{Neutze:2000,Chapman:2011}. The pulse subsequently destroys the exposed
crystal so only a single exposure is possible from each. Serial crystallography is also
carried out at synchrotron radiation facilities, to acquire data at exposures far below
radiation damage limits, with the only limitation being that there is enough signal
measured in each diffraction pattern to permit accurate indexing and prediction of the
locations of Bragg peaks~\cite{Gati:2017}. Obtaining a 3D map of structure factors at a
desired minimum signal to noise ratio is then a matter of measuring a sufficient number of
diffraction patterns, provided that time and sample are available. Serial crystallography
is suitable for time-resolved crystallography of dynamical systems triggered by light or
by mixing of a reagent, particularly since exposure times are short and there is no need
to restore the crystal to the ground state. Following recent developments to analyse
diffraction patterns that contain only several Bragg peaks~\cite{Li:2019b,Stockler:2023},
it has recently been applied to the measurement of small-molecule
crystals~\cite{Schriber:2022,Takaba:2023,Kang:2024,Moon:2024}.

Snapshot diffraction patterns of well-ordered crystals made using a collimated
quasi-monochromatic X-ray beam consist mainly of partial reflections. This requires tens
of thousands of patterns to be measured to obtain accurate estimates of structure
factors~\cite{White:2013}. One way to speed up serial crystallography is to record
predominantly fully-integrated reflections by utilising the full ``pink beam'' spectrum of
an X-ray undulator~\cite{Meents:2017}. This is the serial-crystallographic extension of
Laue diffraction. A Laue diffraction pattern (that is, the still diffraction pattern of a
three-dimensional crystal obtained with a broad bandwidth) consists of many more
reflections than does a monochromatic pattern since each peak in the diffraction pattern
is formed by a particular wavelength that obeys the Bragg condition for a given spatial
frequency or reciprocal lattice vector of the crystal. Laue diffraction was successfully
used for early time-resolved macromolecular crystallography
experiments~\cite{Moffat:1984,Hajdu:1987}, where the increase in total fluence compared
with a monochromatic beam enabled exposure times at a synchrotron radiation facility as
short as about \SI{100}{\pico\second}~\cite{Schotte:2003}, and the increase in the
coverage of reciprocal space for a single orientation of the crystal allowed full datasets
to be collected from just a few (comparatively large) crystals and crystal
orientations. Assuming known lattice parameters, indexing the diffraction pattern entails
determining both the Miller index corresponding to each reflection and the particular
wavelength. This is a considerably more challenging problem than indexing a monochromatic
pattern, especially for snapshot patterns in serial crystallography, since the assignment
of each spot to a reciprocal lattice vector depends both on the scattering angle and the
unknown wavelength. The pinkIndexer program indexes Laue patterns by finding a consensus
solution from many peaks~\cite{Gevorkov:2020}. In trials of pink-beam serial
crystallography, the full 2\% bandwidth of an undulator source at a synchrotron radiation
facility resulted in structure determination from only 50 still
patterns~\cite{Meents:2017}. A four-fold reduction over monochromatic data collection was
achieved with XFEL pulses of 2\% bandwidth~\cite{Nass:2021}. The lower reduction in the
number of patterns compared with using synchrotron radiation may be due to the fluctuating
spectrum of XFEL pulses, reducing the precision of structure factor estimates. If the
spectrum is measured on each pulse and the wavelength corresponding to each peak can be
identified, then it may be possible to correct for the spectral weighting of each
diffraction peak, but this has not yet been demonstrated. It should also be noted that as
the bandwidth increases, the signal to background ratio of peaks is reduced since only a
small spectral component contributes to each peak, whereas the background is proportional
to the total flux.

A particularly elegant approach for time-resolved data collection, put forth by
Moffat~\cite{Moffat:2003}, is to carry out Laue diffraction with chirped X-ray pulses.
In a chirped pulse, the instantaneous wavelength varies with time. Identifying the
wavelength associated with each Bragg peak, by indexing the pattern, allows diffraction intensities to
be assigned to a particular time of occurrence, $T_X$. Many measurements, such as those in serial
crystallography, could then be aggregated to obtain a complete set of structure factors as
a function of delay $T = T_X-T_p$ relative to the arrival of an optical pump pulse
at $T_p$, for example. The time
resolution in any such pump-probe experiment is given by the quadrature sum of the
duration of the pump pulse, the duration of the probe pulse, and the timing jitter between
them. A chirped pulse can potentially reduce the contribution of jitter to this timing
error by encoding time in the observable diffraction pattern, and allowing the
durations spanned by different measurements to overlap so they can be registered to each
other. Since the chirped pulses are longer than the short-pulse unchirped pulses needed to
achieve a certain time resolution, the chirped pulses may be of lower intensity and thus
less destructive~\cite{Moffat:2003}.

\section{Convergent-beam diffraction}
\label{sec:CB-diff}
Given that the Bragg equation $\lambda = 2 d \sin \theta$ has only three variables---the
wavelength $\lambda$, the period of the crystal planes $d$, and the Bragg angle of the
reflection $\theta$---an alternative to Laue diffraction is to supply a range of angles of
incidence that illuminate the crystal. This is the case in
convergent-beam crystallography, where the beam is brought to a tight focus by a lens of
high numerical aperture (NA) such as a multilayer Laue lens~\cite{Bajt:2018} and used to
illuminate the sample. Such lenses
can now focus to spots that are as small as the unit-cell dimensions of protein
crystals~\cite{Dresselhaus:2024}. Equivalently, the angular extent of the focused beam can
exceed the angular separation of Bragg peaks. The convergent illumination vastly increases
the number of recorded reflections in a single monochromatic diffraction pattern and
ensures that they are fully integrated. And just as Laue diffraction offers a way to
encode time using a chirped pulse~\cite{Moffat:2003,Fadini:2020}, we show that time can also be encoded
in convergent-beam diffraction patterns with a remarkably high resolution, using two
distinct schemes. The approaches are compatible with the spectral properties of X-ray FEL
pulses, including newly-developed attosecond pulses.  

\begin{figure*}[tb]
  \centering
  \includegraphics[width=\linewidth]{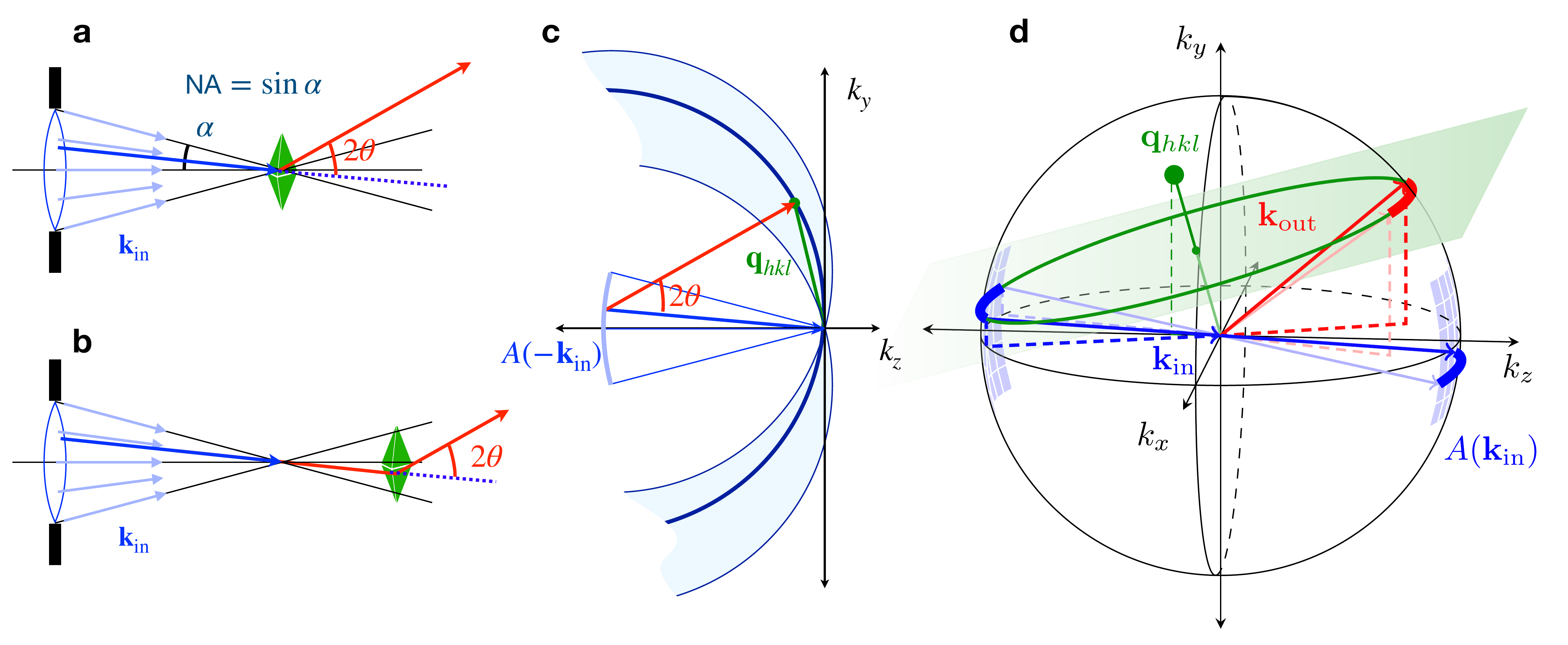}
  \caption[Convergent-beam diffraction]{Convergent beam diffraction from a thick crystal
    in focus (a) occurs within the volume of the crystal intersected by the focused
    beam. When out of focus (b), diffraction occurs at the positions intersected by the particular $\kin$
    vectors (blue) that fulfil the diffracting condition. (c) In both cases, the diffracting condition is determined
    by the orientation of the Ewald sphere intersecting the reciprocal lattice vector
    $\vec{q}_{hkl}$. (d) The condition is invariant to rotation around the $\vec{q}_{hkl}$ vector and so the diffracting
    and incident vectors describe a circle in the plane normal to and bisecting $\vec{q}_{hkl}$. Given the
    available $\kin$ supplied by the lens aperture $A(-\kin)$, these vectors are confined to the surface of
    a cone, with its apex at the origin, and sweep out arcs (bold lines). }
  \label{fig:CB-diff}
\end{figure*}

We consider an X-ray beam focused by a lens.  The beam near the focus can be described by
its angular spectrum of plane waves, whose amplitude and phase are given by the lens pupil
function $A(-\kin)$, a complex-valued function of wave-vectors $\kin$ that all have
magnitude $1/\lambda$. (We use the crystallographic convention that
$|\mathbf{k}| = 1/\lambda$ and that the magnitude of a reciprocal space vector
$|\mathbf{q}|=1/d$, for a spatial period $d$.) The real-space distribution of the field in
the focus, $a(\vec{x})$, is the synthesis of all the plane wave components, given by the
Fourier transform of $A$. Furthermore, the distribution at small distances $z$ from the
focus can be related to the Fresnel transform of $a(\vec{x})$.  In the far-field of the
focus, which can be reached in just a few hundred Rayleigh lengths, the distribution of
the wave-field is again $A(\kin)$. This is certainly the case at the plane of the detector
where the intensity of the beam diverging from the focus is proportional to $|A(\kin)|^2$.
The sine of the semi-angle $\alpha$ of the angular distribution of the wave-vectors
supplied by the lens is referred to the numerical aperture of the lens, or NA, shown in
Fig.~\ref{fig:CB-diff} (a).

The formalism of convergent-beam diffraction is often presented within the context of thin
objects that are described by a two-dimensional transmission function, such as in the
analysis of ptychography where the object is placed in the focal plane and diffraction is
recorded as the objects is scanned in this transverse plane~\cite{Rodenburg:2019}. The
diffraction pattern for each position of the object is given by the Fourier transform of
the product of the transmission function with the probe function $a(\vec{x})$. For a
two-dimensional crystal, for example, the diffraction therefore consists of the entire
pupil function, $A$, convolved with each diffraction order. If the angular separation of
the diffraction orders are close enough then the pupil functions will
overlap and, in the regions of overlap, will thereby interfere, giving the opportunity to
determine the relative phases of those orders~\cite{Spence:2014b}.
This occurs if the diameter of the pupil, $2 \mathrm{NA}$,
exceeds the angular separation $\lambda/d$ of the orders, which is equivalent to the
condition that the width of the focus formed by the unaberrated lens,
$\lambda/(2 \mathrm{NA})$, is smaller than the period $d$, stating that that period can be
resolved by the lens.

Here, however, we consider a \emph{three-dimensional} object placed at or near the focus,
and determine the diffraction conditions utilising the construction of the Ewald sphere
and the 3D Fourier spectrum of the object. The Ewald sphere is the two-dimensional
manifold in 3D reciprocal space that intersects the spatial frequencies $\vec{q}$ (of the
illuminated object) that diffract an incident monochromatic plane wave with wave-vector
$\kin$. The Ewald sphere intercepts the origin, $\vec{q} = 0$, is centered at $-\kin$, and
has a radius $1/\lambda$. Since the focused beam incident on the diffracting object can be
described as a coherent sum of plane wave components supplied by the lens, the diffraction
pattern can be computed as the coherent sum of the diffraction from each component as
determined by the intersection of the corresponding Ewald sphere with the object’s Fourier
spectrum, and after applying the weights $A(\kin)$. The complement of Ewald spheres from
all wave-vectors provided by the lens fills a volume of reciprocal space, illustrated in
Fig.~\ref{fig:CB-diff} (c), that contributes to the convergent-beam diffraction pattern.

\begin{figure}[tb!]
  \centering
  \includegraphics[width=6cm]{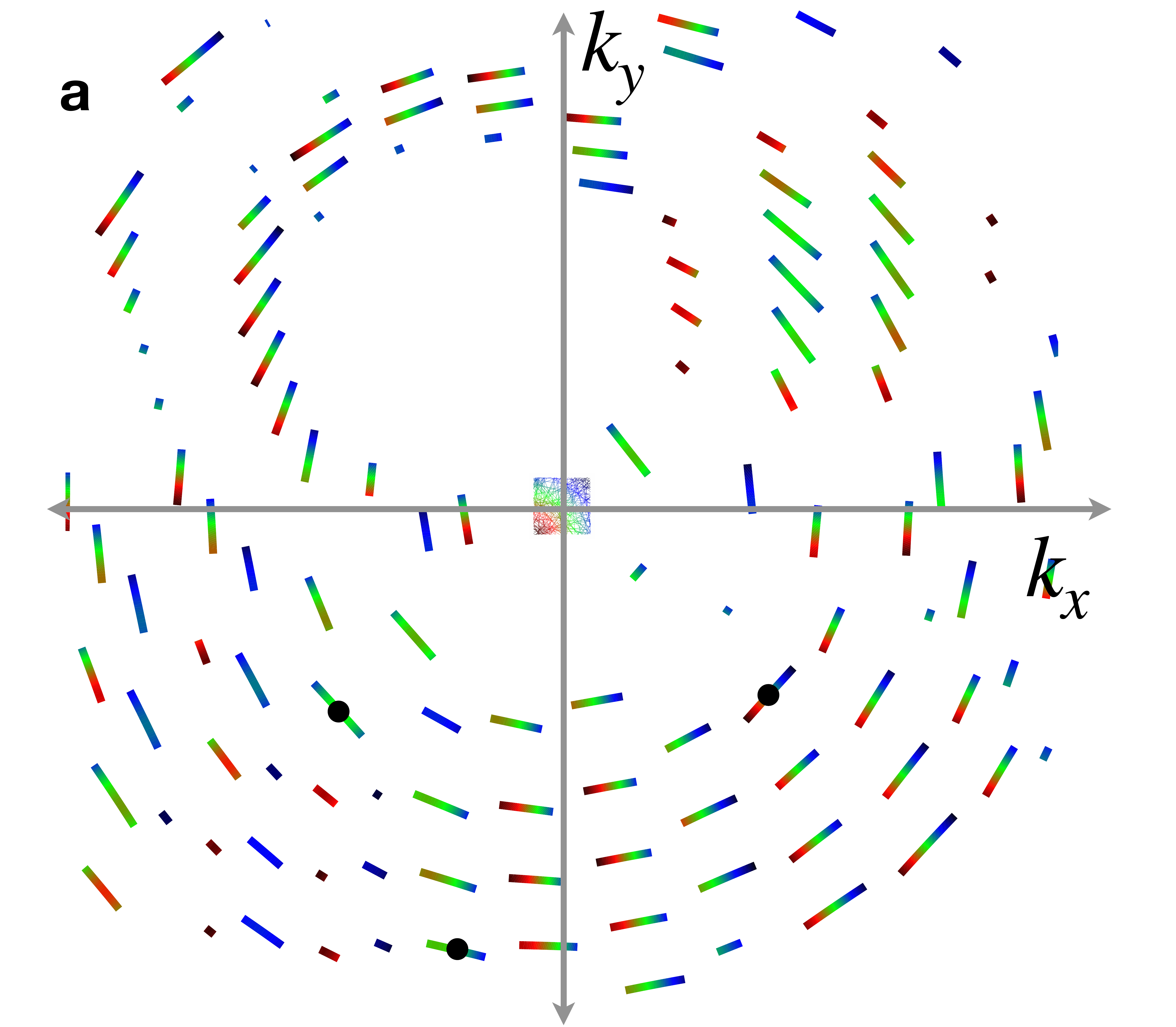}
  \includegraphics[width=6cm]{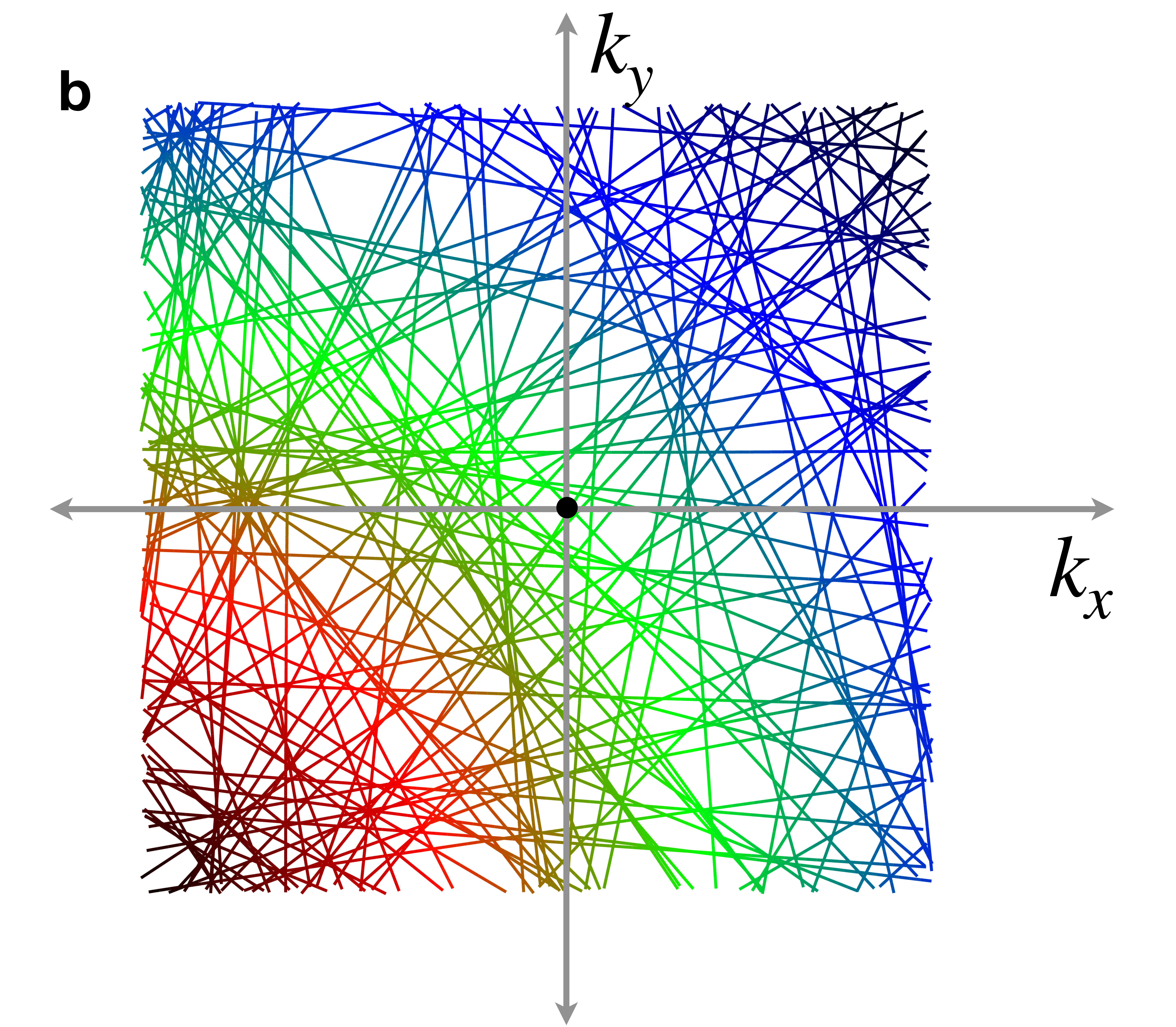}
  \includegraphics[width=6cm]{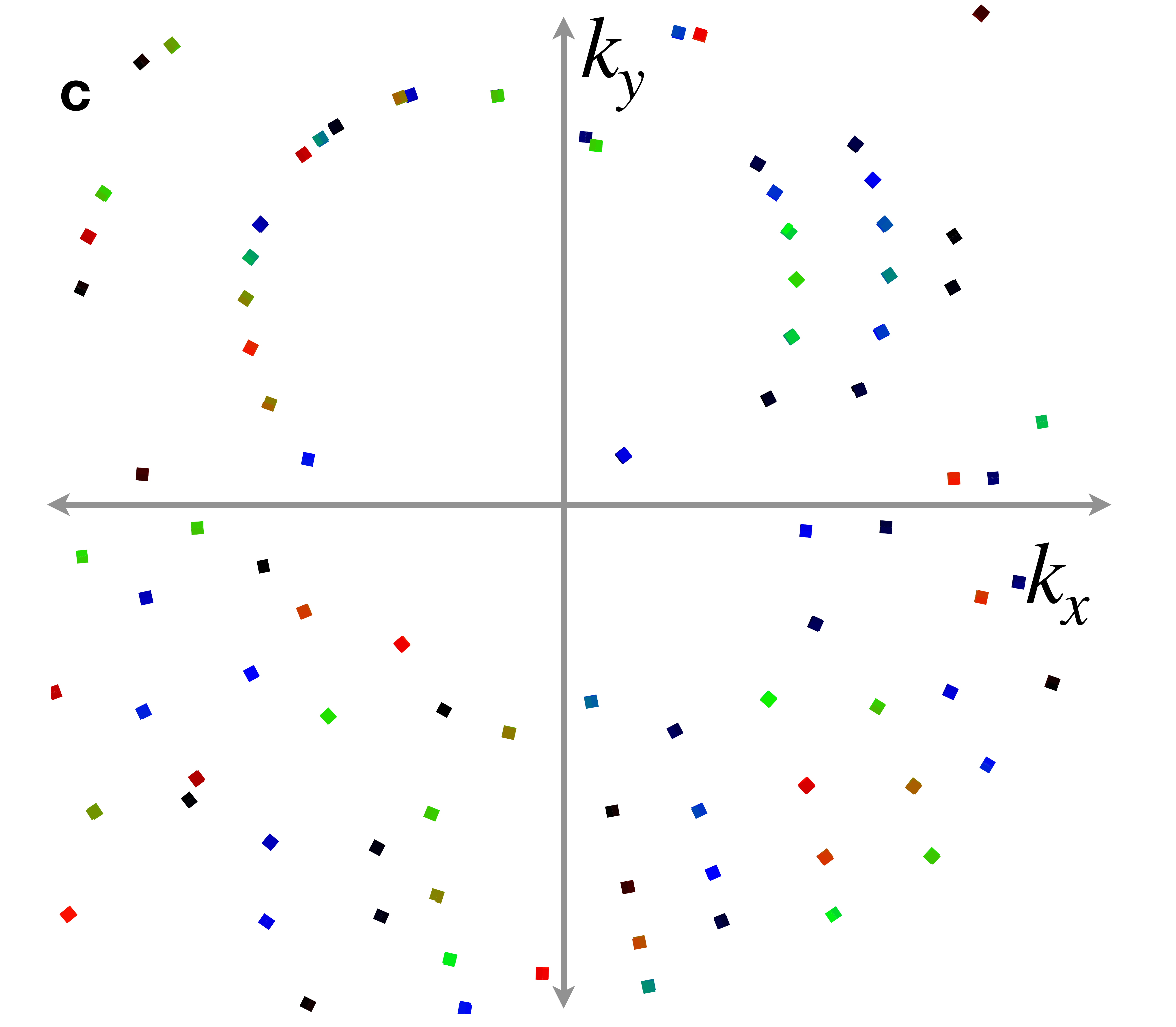}
  \caption[Diffraction patterns]{(a) Simulated convergent-beam diffraction pattern showing
    positions of Bragg streaks to a resolution of \SI{1.5}{\angstrom} (in the corners of
    the pattern) at a wavelength of \SI{1}{\angstrom} with an off-axis lens with a square aperture,
    NA = 0.035, and for an orthorhombic lattice with unit cell parameters
    $a = \SI{9.0}{\angstrom}$, $b= \SI{15.7}{\angstrom}$, $c= \SI{18.8}{\angstrom}$. The
    center of the pattern shows the transmitted deficit lines that map out the lens
    pupil and shown magnified in (b).  The deficit lines and Bragg streaks are coloured
    according to the transverse distance in the lens pupil from the bottom left
    corner. Black circles indicate the positions of Bragg peaks that would be seen if the
    NA was reduced to 0.001. (c) Laue diffraction pattern obtained from the same crystal
    lattice with a relative bandwidth of 25\%. In this case the colour represents
    wavelength, ranging from \SI{0.87}{\angstrom} (dark blue) to \SI{1.12}{\angstrom}
    (dark red).  }
  \label{fig:pattern}
\end{figure}

For a crystalline object, each reciprocal lattice vector $\vec{q}_{hkl}$ within this
volume selects an incident $\kin$ according to which Ewald sphere it intersects. The
wave-vector of the diffracted beam is given by $\kout = \vec{q}_{hkl} + \kin$. Since the
range of $\kin$ is two dimensional, there is not just one unique incident wave-vector that
satisfies the Bragg condition for a particular reciprocal lattice point as would be the
case for Laue diffraction. Indeed, all wave-vectors supplied by the lens that can be
formed by rotating $\kin$ around the $\vec{q}_{hkl}$ vector will also obey that condition,
as seen in Fig.~\ref{fig:CB-diff} (d). Since the magnitudes of $\kin$ and $\kout$ are equal, their average
is perpendicular to and bisects $\vec{q}_{hkl}$ . The rotation of $\kin$ around
$\vec{q}_{hkl}$ sweeps out a circle in the plane normal to $\vec{q}_{hkl}$ and which
intersects the point $\vec{q}_{hkl}/2$. The circle (shown in green in Fig.~\ref{fig:CB-diff} (d)) can be
found from the solution of the two equations
\begin{equation}
  \label{eq:1}
  \mathbf{k}^2 = \frac{1}{\lambda^2}; \:\:\:\: \left(\mathbf{k}-\frac{\mathbf{q}_{hkl}}{2}\right)\cdot \mathbf{q}_{hkl}=0
\end{equation}
This circle has radius equal to the magnitude $\bar{\mathbf{k}}=(1/\lambda) \, \cos
\theta$
of the average vector $\bar{\mathbf{k}} = (\kin+\kout)/2$, and can be
parameterised as
\begin{equation}
  \label{eq:2}
  \mathbf{k}_p(\chi)=\frac{\mathbf{q}_{hkl}}{2}+\cos \chi\,\bar{\mathbf{k}}_0+\sin\chi\,\bar{\mathbf{k}}_\perp
\end{equation}
where $\bar{\mathbf{k}}_0$ is initially solved in the plane containing $\mathbf{q}_{hkl}$
and the optical axis (the $z$ axis) and $\bar{\mathbf{k}}_\perp$ is set orthogonal to both
$\bar{\mathbf{k}}_0$ and $\mathbf{q}_{hkl}$ from their cross product. The range of $\chi$
for the diffracted $\kout(\chi)=\mathbf{k}_p(\chi)$ and the incident
$\kin(\chi)=\mathbf{k}_p(\pi-\chi)$ is limited by the extent of the lens pupil. The diffraction
due to the reciprocal lattice vector therefore consists not of a Bragg peak but a line,
which we refer to as a Bragg streak. On a flat detector this is a curved line given by the
intersection of the cone of $\kout(\chi)$ with the plane of the detector, but for a small
NA the streaks are well approximated by straight lines. The simulated pattern
in Fig.~2 (a) shows the positions of Bragg streaks for an orthorhombic reciprocal lattice
with unit cell dimensions of $a=\SI{9.0}{\angstrom}$, $b = \SI{15.7}{\angstrom}$,
$c = \SI{18.8}{\angstrom}$, chosen to be comparable to those of a pharmaceutical
compound. The crystal was taken to be in a non-special orientation, and the pattern was
simulated for a lens with $\mathrm{NA} = 0.035$, and a wavelength of \SI{1}{\angstrom}.
For this particular
orientation there are 139 Bragg streaks out to a resolution of \SI{1.5}{\angstrom}.

Each $\kout(\chi)$ Bragg streak in Fig.~\ref{fig:pattern} (a) has a corresponding
$\kin(\chi)$ “deficit line” that cuts through the pupil as shown in the center of the
pattern and magnified in Fig.~\ref{fig:pattern} (b). Each deficit line is the line of
incident wave-vectors that are diffracted into the corresponding Bragg streak, and thus
has the same orientation and length as the Bragg streak. The naming is due to the fact
that in thick crystals, described by dynamical diffraction theory, the transfer of
intensity into the Bragg streak is seen as a loss of intensity in the forward beam, giving
a dark line across the direct beam (the pupil function) projected on the
detector~\cite{Spence:1992}. Although for small protein crystals, this loss of intensity
is likely not observable, the position of the $\kin(\chi)$ deficit line is nevertheless
important because it may map to particular positions across the face of the crystal or
delay times, as described below in Secs.~\ref{sec:out-of-focus} and \ref{sec:TR}. The
deficit lines can be determined by indexing the pattern, and indeed the constraint that
the $\kin(\chi)$ lines are bounded by the lens pupil can be used to refine the indexing
solution (Chufeng Li, in preparation).  In Fig.~\ref{fig:pattern} (b), the $\kin(\chi)$
lines are coloured according to the distance from the bottom left corner, and the
corresponding $\kout(\chi)$ Bragg streaks are given the same colour modulation in
Fig.~\ref{fig:pattern} (a). Thus, streaks that are oriented diagonally from bottom left to
top right change colour from red to green to blue whereas streaks in directions orthogonal
to that tend to have a single colour.

Looking at Fig.~\ref{fig:pattern} (b), and considering that the Bragg lines lie on full
circles $\mathbf{k}_p(\chi)$ as depicted in Fig.~\ref{fig:CB-diff} (d), one can appreciate
that the number of participating reflections increases with the numerical aperture of the
lens. Clearly, if we stop down the aperture, some of the streaks will no longer be
included. Reducing the NA to zero, which is the case of a collimated incident beam, would
likely only intersect a small number of reflections for this particular lattice, if any,
depending on how many deficit lines are intersected, and even then these may only be
partial reflections. The gaps between deficit lines indicate that, at least for a perfect
crystal, there are many orientations of the crystal for which no lines would be
intersected. For a quasi-collimated incident beam with a divergence of
\SI{1}{\milli\radian} only three reflections will be observed for this particular case as
shown in Figs. 2 (a) and (b) by the black circles. (An equivalent result would occur for a
crystal with \SI{1}{\milli\radian} mosaicity illuminated by a collimated incident beam.)

Figure \ref{fig:pattern} (b) also highlights the fact that in convergent-beam diffraction, the Bragg
streaks predominantly provide fully integrated intensities across the short width of the
Bragg streak (i.e. in the direction of changing Bragg angle). The total counts along the
entire length of the Bragg streak depends on its intersection with, and weighting by, the
pupil intensity function $|A(\mathbf{\kin})|^2$, and for uniform illumination the counts can be simply
normalised by the length of the streak. If the crystal is placed out of focus as depicted
by Fig.~\ref{fig:CB-diff} (b), then the Bragg streak intensity is also modulated by the spatial profile of
the diffraction efficiency of the crystal as discussed below in Sec.~\ref{sec:out-of-focus}.

\begin{figure*}[tb!]
  \centering
  {\fontfamily{phv}\selectfont
  \begin{overpic}[width=0.8\linewidth]{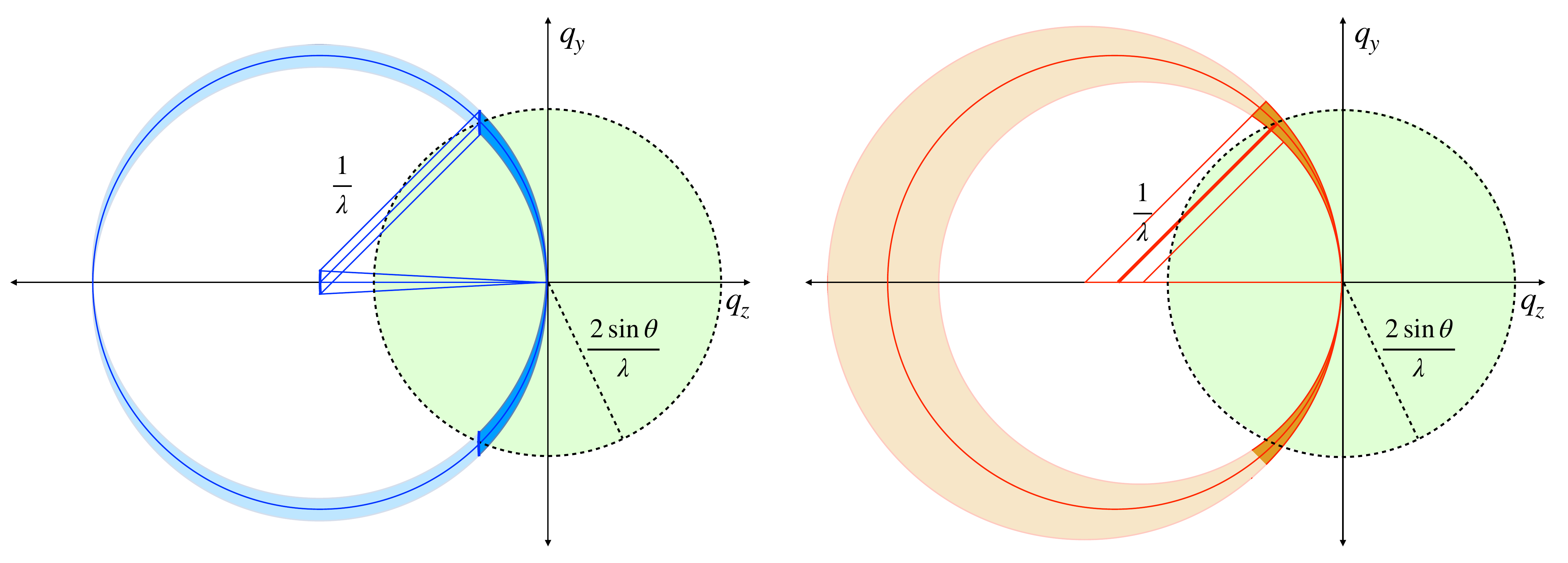}
    \put(2,32){\textbf{a}}
    \put(52,32){\textbf{b}}
  \end{overpic}
  \begin{overpic}[width=0.8\linewidth]{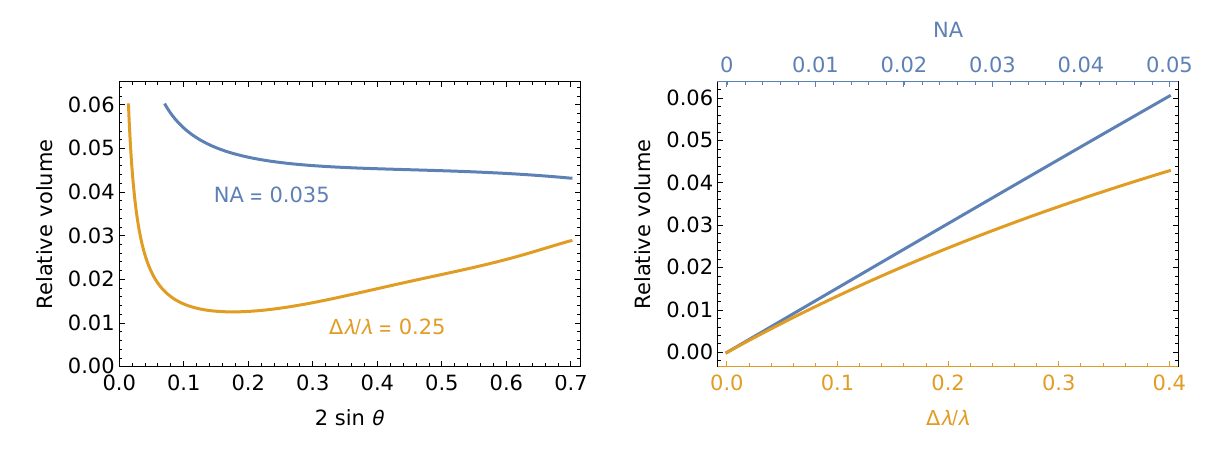}
    \put(2,32){\textbf{c}}
    \put(52,32){\textbf{d}}
  \end{overpic}
  }
  \caption[Ewald volumes]{Convergent-beam diffraction patterns intersect large volumes of
    reciprocal space, comparable to high-bandwidth Laue diffraction. (a) The participating
    volume of reciprocal space in convergent-beam diffraction lies between Ewald spheres
    centered on the range of $-\kin$ vectors, shown shaded in blue. When considering
    diffraction extending to a maximum angle $2\theta$, only the region shaded in
    dark blue participates. This volume is a fraction of the full volume of
    resolution-limited reciprocal space (indicated by the green shading), referred to as
    the relative volume. (b) Similarly, for Laue diffraction, the dark orange volume gives
    the participating volume of reciprocal space for a given maximum scattering angle. (c)
    The relative volumes are plotted as a function of
    the resolution for $\mathrm{NA} = 0.035$ for convergent-beam diffraction (blue) and for
    a relative bandwidth of 25\% for Laue diffraction (orange). (d) The relative volumes
    for $2\theta = \SI{45}{\degree}$ ($2\sin\theta =0.7$), plotted as a function of the
    lens NA or relative bandwidth.}
  \label{fig:Ewald}
\end{figure*}

As mentioned above, Laue diffraction, which employs a collimated incident beam of broad
bandwidth, also gives rise to many Bragg reflections. The pattern of these reflections is
given in Fig.~\ref{fig:pattern} (c) for the same crystal and orientation as for the
convergent-beam diffraction example in Fig.~\ref{fig:pattern} (a). In this case the
bandwidth was taken to be 25\%, centered at a wavelength of \SI{1}{\angstrom}, for which
there are 218 potential reflections to a resolution of \SI{1.5}{\angstrom}. As with
convergent-beam diffraction, the reflections are fully integrated and the associated
central wavelength for each reflection is only found after indexing. The intensities are
modulated by the spectrum of the incident beam. It is perhaps surprising that the number
of reflections in the convergent-beam diffraction pattern with $\mathrm{NA} = 0.035$ is
comparable to the number of reflections in the Laue diffraction pattern with a bandwidth
of 25\%. Assuming that the density of reciprocal lattice vectors is relatively homogeneous
throughout reciprocal space, the number of Bragg peaks or streaks in a pattern should be
proportional to the reciprocal-space volume enclosed by the limiting Ewald spheres of
incident beam angles or wavelengths. The volume addressed by a 25\% bandwidth is much
greater than the volume addressed by a numerical aperture of 0.035, but this is only the
case when the entire resolution limit is considered. The derivative of Bragg’s equation
shows that the equivalence of relative bandwidth and convergence angle depends on the
Bragg angle according to
\begin{equation}
  \label{eq:3}
  \frac{\Delta \lambda}{\lambda}=\frac{\Delta \theta}{\tan \theta }
\end{equation}
and so at low resolution, a small convergence angle is equivalent to a larger relative
bandwidth than at a higher resolution. Figures~\ref{fig:Ewald} (a) and (b) show a cut
through the $k_y$ - $k_z$ plane showing that when the comparison is limited to reflections
at scattering angles $2\theta < \SI{45}{\degree}$, or resolutions $1/d < 0.77/\lambda$,
the volumes between the limiting spheres do appear comparable. Clearly, Laue diffraction
provides a large participating Ewald volume at high resolutions in the back-scattering
direction. The participating volumes for reflection are compared with the total reciprocal
space volume for a given maximum scattering angle in Fig.~\ref{fig:Ewald} (c) for
convergent-beam diffraction and Laue diffraction. Here, the relative volume is defined to
be equal to the ratio of the participating volume divided by the volume of a sphere of
radius $(2\sin\theta)/\lambda$. As such, for a given resolution, it is the fraction of
reciprocal space that can be encountered in a single exposure. It is thus the reciprocal
of the minimum possible number of exposures to measure a complete dataset at that
resolution, assuming P1 crystal symmetry. Fig.~\ref{fig:Ewald} (d) plots the relative
volume as a function of the numerical aperture or relative bandwidth for the two methods
at a maximum scattering angle of $2\theta = \SI{45}{\degree}$.

For completeness, a diffraction pattern recorded with a broad bandwidth convergent beam
will consist of Bragg patches rather than streaks. If the wavelength is varied, the Bragg
angle of each Bragg streak in a monochromatic convergent beam diffraction pattern changes
according to Bragg’s law, sweeping the Bragg streaks in the radial direction of the
diffraction pattern and moving the $\kin(\chi)$ deficit line across the pupil in a direction perpendicular
to the line. Diffraction will occur as long as the deficit line falls within
the lens pupil, so for a broad enough bandwidth each Bragg patch will be a mapping of that
pupil and could provide a full topogram of a crystal placed out of focus (see
Sec.~\ref{sec:out-of-focus}). These Bragg patches may overlap, in which case they will tend to add
incoherently since a given detector pixel would receive wave-vectors of differing
wavelength.

\subsection{In-focus diffraction}
\label{sec:in-focus}
When the crystal is placed in focus then all reflections originate from the same volume of
the crystal, given by the overlap of the three-dimensional focal distribution with the
volume of the crystal. Even if the focal spot is smaller than a unit cell of the crystal
the diffraction pattern will be indicative of the periodicity since the transverse
component of the focal distribution $a(\vec{x})$ is not strictly truncated. The focal spot
for a perfect lens with a square pupil, for example, is described by a sinc function, and
each plane wave component of the beam extends beyond the width of the focus. If the
crystal is sufficiently thick then the pattern will consist of Bragg streaks as depicted
in Fig.~\ref{fig:pattern} (a). This is due to the periodicity of the crystal structure
along the depth of the focus, which is visible when viewed back from the direction of the
outgoing Bragg reflection~\cite{Hruszkewycz:2017}. As the analysis above shows, the width
of the Bragg streak (in the direction of increasing scattering angle $2\theta$) is
proportional to the width of the reciprocal lattice peak in the $k_z$ direction, which
itself is inversely proportional to the thickness of the crystal as seen in the projected
view of the reflected beam. The intensity profile in this direction is the rocking curve
of the reflection and so can be integrated to give a direct estimate of the structure
factor, after first normalising by any variation in the pupil function,
$|A(\kin)|^2$. This first requires knowing the pupil coordinates of the $\kin(\chi)$ vectors
that gave rise to each Bragg streak, which itself requires indexing the pattern to solve
for the crystal orientation. An approach for indexing such patterns was addressed by
Gevorkov~\cite{Gevorkov:2020}. Since the strength of diffraction is given by the product of
the beam intensity (photons per area) and crystal volume, illuminating such a small volume
of the crystal (within the focus) would demand a dose that is far beyond damage limits of
a radiation-sensitive sample such as a macromolecular crystal. In-focus measurements of
such samples are only practical by out-running damage with an X-ray free-electron laser
pulse.

\subsection{Out-of-focus diffraction}
\label{sec:out-of-focus}
A larger volume of the crystal can be exposed at lower dose by moving it out of the beam
focus. In the transverse plane at a defocus distance of many Rayleigh lengths (in the far-field of
the focus) the beam is predominantly a diverging (or converging) spherical wave for which
the ray direction $\kin$ is directly correlated to its transverse position as
\begin{eqnarray}
  \label{eq:4}
  \begin{aligned}
  (x,y) &= \frac{(k_x,k_y)\,z}{\sqrt{1/\lambda^2-k_x^2-k_y^2}} \nonumber \\
  &=  (\tan \phi_x,\tan
  \phi_y)\,z \approx (\phi_x, \phi_y)\,z
  \end{aligned}
\end{eqnarray}
A crystal placed in this plane produces Bragg streaks that can therefore be mapped back to
illuminating $\kin(\chi)$ lines that cut across this plane and hence across the face of
the crystal as shown in Fig.~\ref{fig:CB-diff} (b). The intensity along a Bragg streak
will be modulated by the diffraction efficiency of the crystal where it is intersected by
the corresponding $\kin$ rays. Obviously, diffraction will not occur beyond the extent of
the crystal. In a non-destructive experiment, rocking or translating the crystal will
cause the intersection lines to sweep across the face of the crystal. A magnified
diffraction topogram of the crystal can therefore be constructed from each Bragg streak,
giving spatial maps of the diffraction efficiency of the particular Bragg reflections
(Chufeng Li, in preparation). The magnification of the topogram is given by the ratio of
the distance of the detector to the focus, divided by the defocus distance of the
crystal. For defocus distances of about \SI{10}{\micro\meter} and a detector distance of
\SI{10}{\centi\meter} the magnification can exceed \num{10000}, such that a
\SI{50}{\micro\meter} wide detector pixel maps to an effective pixel size at the crystal
of only \SI{5}{\nano\meter}.  As with projection imaging of non-periodic objects with
high-NA lenses~\cite{Zhang:2024}, the magnified topogram equivalent to the defocused image
that would be obtained of the crystal using a lens that forms a (defocused) image directly
on a detector. In this case, the equivalence is to a dark-field
image~\cite{Simons:2015}. The defocus gives phase contrast that may reveal lattice defects
and the image can be simulated by the Fresnel transform of the projected diffraction
efficiency of the crystal (projected along the direction of the incident beam).

Examining Fig.~\ref{fig:CB-diff} (d) shows it is necessary to rotate the crystal by an
angle of at least $2\alpha$ to sweep $\kin(\chi)$ across the full pupil, for a rotation
axis that is perpendicular both to $\vec{q}_{hkl}$ and the beam direction (e.g.\ around
the $k_x$ axis in Fig.~\ref{fig:CB-diff} (d)). Bragg streaks that are oriented parallel to
the $k_x$ axis (such as the horizontal streaks in Fig.~\ref{fig:pattern} (a)) would be
swept in the direction of $k_y$. Bragg streaks located away from the $k_y$ axis may
require larger rotation ranges to effect a full sweep, and streaks on the $k_x$ axis will
not change position with a rotation about that axis. However, it can be seen that full
topograms from a large number of Bragg reflections could be obtained from only a small
rotation about a single axis.

A magnified topogram of the diffraction efficiency can also be obtained from a single
snapshot pattern, albeit constructed from all available Bragg streaks in the pattern that
are mapped back to their $\kin(\chi)$ positions (solved by indexing), such as shown in
Fig.~\ref{fig:pattern} (b). The intensity profile of each Bragg streak is given by the
product of the pupil function $|A(\kin)|^2$, the two-dimensional map $C(x,y)$ of the
projected diffraction efficiency of the crystal, and the square modulus
$|F(\mathbf{q}_{hkl})|^2$ of the structure factor of that reflection. Obtaining precise
estimates of the structure factors from a single pattern thus requires normalising by the
pupil function and by $C(x,y)$. 
Since the pupil function is known (from the direct beam, measured without a beamstop) it
is possible to solve for $C(x,y)$ and the structure factors. This is because, to a good
approximation, the crystal contribution is common for all reflections that intersect a
particular $(x,y)$ position and the structure factor contribution is common for the entire
Bragg line. The effect of any variation of the crystal diffraction efficiency can be
thought of as being similar to a partiality factor, in that each reflection is generated
by a different partial volume of the crystal. Solving for this partiality is a matter of
finding self-consistent set of scalings $C(x,y)$ and no additional model of the crystal
structure is necessarily required (at least, not for static structure factors). However,
the assumption that the crystal has uniform thickness and diffraction efficiency would
imply the counts in each Bragg streak can be simply normalised by its length.

\section{Time-Resolved Convergent Beam Diffraction}
\label{sec:TR}
As mentioned above in Sec.~\ref{sec:serial}, time can be encoded in a Laue diffraction pattern of a crystal
using a chirped pulse, in which the wavelength varies with time~\cite{Moffat:2003}. Since each
Bragg peak in a Laue diffraction pattern selects a particular wavelength of the spectrum,
with a chirped pulse each peak will therefore correspond to a particular arrival time on
the crystal of the particular spectral component. That time can be determined by indexing
the pattern, thereby solving for the wavelength of each Bragg peak. A particular
volume of the crystal is exposed throughout the course of the exposure, so
measurements at later times are from a structure that has received the prior cumulative
exposure. Keith Moffat also suggested the use of monochromatic angularly chirped pulses, in which
now the arrival time varies with the angle of incidence~\cite{Moffat:2003}. Given that the indexing of a
convergent-beam diffraction pattern determines the range of incident wave-vectors for each
Bragg streak, it should also be possible to encode and decode time in this case as
well. While creating wavelength-chirped pulses requires particular operation of the XFEL
facility or dedicated instrumentation \cite{Li:2022a}, an angularly chirped pulse is produced
naturally from a monochromatic pulse by using dispersive focusing optics. In the X-ray
regime, these include either refractive or diffractive lenses. 

\begin{figure*}[tb!]
  \centering
  \includegraphics[width=0.8\linewidth]{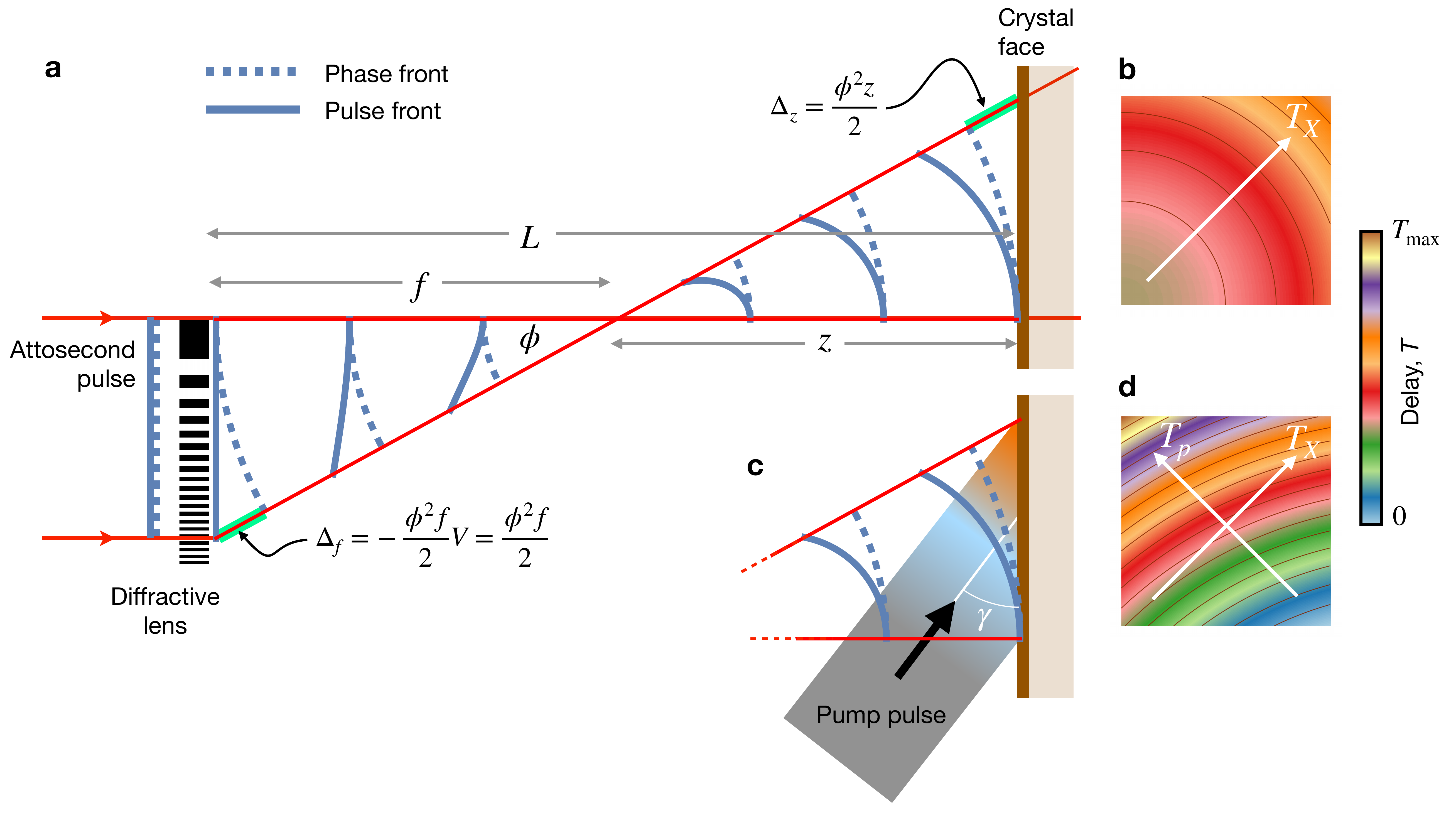}
  \caption[A short pulse]{(a) A short collimated pulse focused by a dispersive lens lags
    behind the phase front by a time $\Delta_f/c$ that depends quadratically on the
    angle $\phi$ of the
    ray with the optical axis, shown here for a diffractive lens with a dispersive power
    of $V=-1$. Rays thus arrive at the focus with this delay. For a flat crystal placed a
    distance $z$ downstream of focus, the rays require an additional time of $\Delta_z/c$ to pass the face
    of the crystal. For a plane-wave pump pulse that arrives simultaneously across the
    crystal face, the pump-probe delay $T$ will vary with arrival time of the probe pulse on the crystal $T_X$,
    given in (b). An inclined pump pulse (c), or one with a tilted pulse front, also maps
    arrival times $T_p$ to position, extending the
    range of delay times to $T = T_X - T_p$ (d).}
  \label{fig:chirp}
\end{figure*}

We usually think of focusing by a lens as being a consequence of Fermat’s principle of
least time: rays all arrive in the focus by travelling along their shortest optical
path. This is the case when considering the phase of the wave. The phase forms a spherical
wavefront that converges to the focus. This wavefront, however, only coincides with the
front of the pulse (or any propagating signal) for a lens without dispersion, which is to
say one whose focal length is independent of wavelength. In a refractive lens, for
example, the refractive index of the material, $n = 1 -\delta$, is slightly less than 1
for X-rays and so the phase velocity in the lens, $c/(1-\delta)$, exceeds the speed of
light in vacuum, $c$. A short pulse propagates through this material at the group
velocity, given by
$v_g = \partial \omega / \partial k = c/(n - \lambda \partial n /\partial \lambda) = c/(1
+ \delta)$, where $\omega$ is the X-ray frequency and we have assumed that the wavelength
is far from absorption edges such that the refractive index decrement $\delta$ is
proportional to $\lambda^2$. Rays that are deflected by the lens at different heights from
the optical axis pass through different thicknesses $l$ of the lens material such that the
pulse front lags behind the wavefront by approximately $2\delta l$. Given that the
thickness of a refractive lens varies quadratically with the transverse height $h$ from
the optical axis (and deflection angle $\phi = h/f$ for a focal length $f$), it is found
that this path difference depends on the square of the deflection angle as $L = \phi^2 f$
~\cite{Bor:1989}. The separation of the pulse front from the wave front for a diffractive
optic such as a multilayer Laue lens has a similar dependence. Starting at the optical
axis, each subsequent period (or bilayer) in a diffractive lens adds one additional
wavelength of path so that rays diffracting toward the focus add constructively. Since the
zone number increases with the square of the ray height, this extra path also increases
quadratically with the lens height. Generally, the pulse front of a beam focused by a
dispersive lens lags behind the phase front by
\begin{equation}
  \label{eq:delta-t}
  T_X(\phi)=-\frac{\phi^2 f}{2 c}\,V
\end{equation}
where
\begin{equation}
  \label{eq:V}
  V = \frac{\lambda}{f}\frac{\partial f}{\partial \lambda}
\end{equation}
is the dispersive power of the lens~\cite{Bor:1988, Chapman:2021b} describing the relative
change in focal length for a relative change in wavelength. For diffractive lenses, the
focal length is inversely proportional to wavelength and so $V=-1$. The quadratic
dependence of $\delta$ on wavelength gives $V=-2$ for refractive lenses, twice the
dispersive power of diffractive optics. As examples, a compound refractive lens stack with
a radius of \SI{500}{\micro\meter} and focal length of \SI{10}{\centi\meter} gives a
maximum propagation time difference of \SI{8}{\femto\second} for rays arriving in the
focus.  Similarly, a typical high-NA multilayer Laue lens may consist of more than \num{10000}
layers, and since two lenses are used together to achieve two-dimensional focusing, the
optical paths of rays reaching the focus exceed \num{20000} waves. At a wavelength of
\SI{1}{\angstrom} (\SI{12}{\kilo\eV} photon energy), this is \SI{2}{\micro\meter} of path
difference, causing the marginal ray to arrive in the focus \SI{6.7}{\femto\second} after
the axial ray.

When a crystal is placed downstream of the focus by a distance $z$, there is an additional
path difference that contributes to the propagation time. As seen from Fig.~\ref{fig:chirp} (a), the
path length of a ray propagating from the focus to the face of the crystal (a transverse
plane), is $z(1/\cos \phi-1)$, so for small angles the angular chirp of the pulse arriving at the plane of
the crystal becomes
\begin{equation}
  \label{eq:delta-t-defocus}
  T_X(\phi)=\frac{\phi^2}{2 c}(z-V\,f).
\end{equation}
Time ranges can therefore be increased by moving a crystal away from focus. Even achromatic
focusing optics such as Kirkpatrick-Baez mirrors, for which $V=0$, provide a temporal variation
across the crystal when it is placed out of focus . The largest time
range of the chirp that can be achieved will be limited by the NA of the lens (setting the maximum
value of $\phi$) and size of the crystal, since
these will limit the range of angles of the diverging beam that intersects the crystal. An advantage
of this scheme is that crystal volumes are not exposed by prior times of the chirp since
rays of differing deflection angle (and hence different delay) intersect different
volumes of the crystal.

A map of the arrival time of the $\kin$ wave-vectors at the crystal face is shown in
Fig.~\ref{fig:chirp} (b) for an off-axis lens, with delay time shown as colour. The
convergent-beam diffraction pattern given in Fig.~\ref{fig:pattern} (a) shows how the
incident $\kin$ rays of Fig.~\ref{fig:pattern} (b) map to the Bragg streaks of a crystal placed in focus
($z=0$). A streak that is oriented with its long axis parallel to the direction of
increasing time in the pupil map thus gives a sweep of time along the Bragg streak. The
structure factor of that particular reflection can be extracted as a function of time by
plotting the intensity along the streak. A streak oriented perpendicular to that will not
have a significant variation in time, but will nevertheless correspond to particular 
arrival times. An on-axis lens, with its pupil centred on the optical axis (where the
arrival time is the shortest), will give rise to Bragg streaks where time increases with
distance in both
directions from some particular point.

A dataset consisting of a full time sweep of the diffraction intensity at
every reciprocal lattice point requires
measuring enough patterns from crystals in different orientations. As compared with the
number of crystals required to obtain a complete dataset irrespective of time, dependent
on the participating volume per exposure plotted in Fig.~\ref{fig:Ewald} (c), a greater
number of exposures are required to obtain completeness at every time point. This can be
crudely estimated by considering how many participating peaks are observed in a
stopped-down pupil that covers a region of where the time is considered
constant (i.e., within a time bin). Given the approximately linear dependence of volume on NA, as seen in
Fig. ~\ref{fig:Ewald} (d), the number of required patterns therefore increases with the
number of time-bins to which the data is apportioned.

\section{Velocity-Matching Conditions}
\label{sec:matching}
The changes in electron density to be measured may be initiated by a pump, such as a pulse
of visible light, UV, or soft X-rays.  Achieving a temporal resolution considerably below
\SI{1}{\femto\second} requires that path lengths of both pump and probe beams are
maintained to much less than \SI{300}{\nano\meter} throughout the entire interaction
volume of the two beams with the molecules in the crystal. This can easily be achieved
when the crystal is thinner than about \SI{300}{\nano\meter} or if the refractive indices
of the crystal are approximately equal for the pump and probe pulses, such as with a soft
X-ray pump. In that case, the pump and probe must be made to propagate colinearly through the crystal volume.

The refractive index for a visible-light or UV pump will differ
considerably from that for the X-ray probe and consequently these pulses will propagate
through the crystal at different speeds. The refractive index for the X-ray probe, $n_X$,
will be approximately equal to unity, whereas the refractive index for the pump, $n_p$,
will likely be larger depending on the wavelength. For example, at visible wavelengths,
the refractive index of a protein crystal may be larger than that of water, which is
1.33. In that case, the pump pulse propagates considerably slower than the probe and would
lag the probe by \SI{1}{\femto\second} after propagating through a sample thickness of
\SI{1.2}{\micro\meter}. Maintaining a fixed delay time along the path of the incident
X-ray beam places constraints on the geometry of the crystal and on the arrangement of the
pump and probe pulses.

\begin{figure*}[tb]
  \centering
  \includegraphics[width=0.85\linewidth]{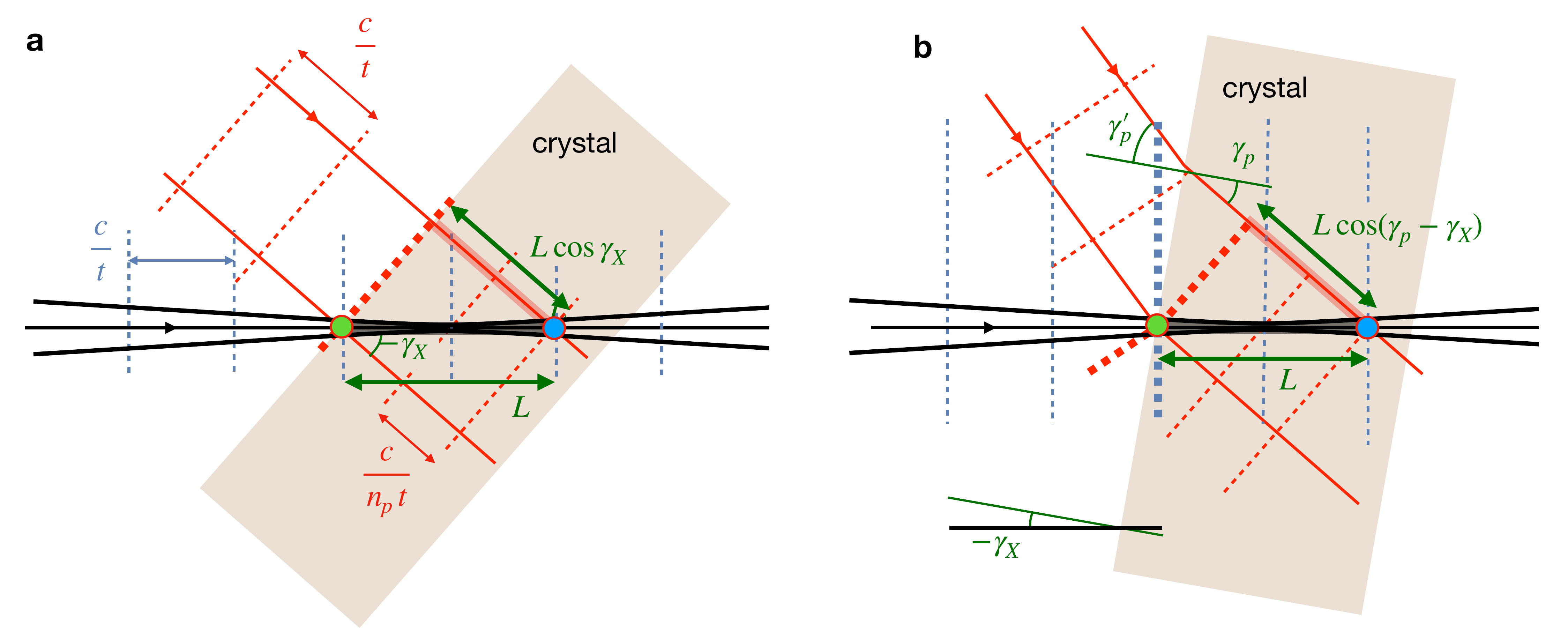}
  \caption[Crystal geometry]{Orientations of the crystal, pump, and probe pulses required
    to maintain the delay between the pump pulse (red) and focused X-ray beam (black),
    along the path of the X-ray beam as it propagates through the crystal with a refractive index $n_p>1$
    for the pump and $n_X = 1$ for the probe. Phase fronts of the X-ray pulse and the pump
    pulse are indicated by blue and red dashed lines, respectively. No pulse tilt is
    assumed. (a) Pump pulse incident normal to the crystal face. (b) Pump pulse refracted
    by the crystal face. }
  \label{fig:crystal}
\end{figure*}

With the crystal in the X-ray focus, this slippage in the case of $n_p > n_X$ can be avoided by using a crystal with
a flat face and tilted so the surface normal is inclined to the X-ray optical axis, as
shown in Fig.~\ref{fig:crystal} (a) where the pump pulse impinges the crystal
perpendicular to its face. The width of the pump pulse should be large enough that it
illuminates the full path of the focused X-ray beam in the crystal. Consider the case
where the X-ray probe and pump pulses arrive simultaneously at the surface of the crystal,
indicated by the green circle in the figure. Given that $n_X \approx 1$, the X-ray beam
then propagates through the crystal medium at a speed $c$ to arrive at a distance $L$
after a time $L/c$, as indicated by the blue circle. This will intersect a different ray
of the pump pulse, travelling with a speed $c/n_p$ but which only has to travel a distance
$L \cos \gamma_X$ in the same time. This is satisfied for all distances $L$ when
$\cos \gamma_X = 1/n_p$, which for $n_p = 1.33$ (as an example) occurs at
$-\gamma_X = \SI{41.2}{\degree}$. Note that when $n_p < n_X$, which may be the case for a
soft X-ray pump, then the geometry must be reversed such that the crystal face is
perpendicular to the X-ray beam and the pump tilted so that it travels a longer path in
the crystal before intersecting the X-ray beam.

Figure \ref{fig:crystal} (a) is not the only available geometry that can satisfy the
velocity-matching condition, which can be achieved over a range of crystal tilts as
illustrated in Fig.~\ref{fig:crystal} (b). In that figure, the pump pulse impinges on the
face of the crystal at an angle and thus refracts, bending the light rays toward the
surface normal. As before, the green circle indicates the point where the X-ray probe and
pump pulses arrive simultaneously at the face of the crystal. At that point in time, the
pulse front of the pump is distributed along the thick dashed line. The ray of the pump
pulse that will ultimately meet up with the probe at the blue circle has at that time
already propagated into the crystal and only need travel a distance
$L \cos (\gamma_p-\gamma_X)$ in the time it takes the X-ray beam to travel a distance
$L$. These two lengths are equal to the case in Fig.~\ref{fig:crystal} (a) (assuming the
same refractive index, $n_p$) and so the relative angle between the pump and the probe
inside the crystal must be the same, requiring that the pump pulse is incident on the
crystal at a steeper relative angle to the X-ray beam, assuming $n_p > 1$. In this case,
the velocity-matching condition is given by
\begin{equation}
  \label{eq:matching}
  \cos (\gamma_p - \gamma_X) = \frac{1}{n_p}
\end{equation}
where $\gamma_p$ is the angle of the refracted pump beam relative to the surface
normal. The incident angle, $\gamma_p'$, is given by Snell's law,
$\sin \gamma_p' = n_p \sin \gamma_p$. As $\gamma_X$ is reduced to zero (with the crystal
face perpendicular to the X-ray beam) the necessary incidence angle of the pump,
$\gamma_p'$, increases to a value given by $\sin \gamma_p' = \sqrt{n_p^2-1}$ with
$\gamma_p' = \SI{61.3}{\degree}$ at $n_p = 1.33$ . The crystal face can continue to be
tilted beyond this angle in the positive sense so that both the pump and probe beams are
incident on the same side of the surface normal until the limit is reached at
$\gamma_p'=\SI{90}{\degree}$, for which $\sin \gamma_p = 1/n_p$. For $n_p = 1.33$ this is
reached at $\gamma_X = \SI{7.5}{\degree}$. 

Since the X-ray beam has some width, there will be a contribution to the error $\Delta T$ in the time
delay $T$ between the pump and probe pulses. For an X-ray beam width of $\delta$ the
difference in delay times between X rays intersecting the same pump ray is given by
\begin{eqnarray}
  \label{eq:5}
  \begin{aligned}
    \Delta T &= \frac{\delta}{c} \left ( \frac{n_p}{\sin (\gamma_p-\gamma_X)} - \frac{1}{\tan (\gamma_p-\gamma_X)}
    \right) \nonumber \\
    & = \frac{\delta}{c}\,  \sqrt{n_p^2-1}
  \end{aligned}
\end{eqnarray}
in the velocity-matching condition of Eqn.~\ref{eq:matching}. The beam width
is greatest out of focus and reaches a maximum of $L \, \mathrm{NA}$ for a crystal
thickness $L$, assuming the beam is focused midway. Thus for a crystal
\SI{3}{\micro\meter} thick, this error reaches \SI{0.3}{\femto\second} for $n_p = 1.33$
and $\mathrm{NA} = 0.035$, giving a mean error over the crystal thickness of
\SI{0.15}{\femto\second}. The error decreases with decreasing $n_p$, NA, or crystal
thickness. Other examples are discussed below in Sec.~\ref{sec:resolution}, following the
analysis of the temporal resolution achievable when the crystal is placed out of focus to
obtain magnified topograms.

The velocity-matching schemes using a crystal with a flat face are required when the refractive
indices of the crystal medium differ significantly for the pump and probe pulses, which is
the case for a visible or UV pump. In this case all the crystals in a serial diffraction
experiment must have faces parallel to each other, located on a larger flat surface that
can be tilted to the X-ray beam and scanned in-plane to expose fresh crystals. Such samples
could be obtained by various means, such as cutting slices of crystals embedded in epoxy
or vitreous ice using a microtome or a focused ion beam, by exfoliation, or by molecular
beam epitaxial growth on a substrate. One promising approach is to grow crystals in a
volume confined by flat parallel membranes, as has been developed for electron
diffraction~\cite{Meents:2023}.

\section{Crossed-Beam Topography}
\label{sec:cross-beam}
While a time-resolved measurement can be made with the crystal placed in focus, using the
angular streaking of $\kin$ wave-vectors mapped from the Bragg-streaks to provide the
temporal range and resolution, the diffraction entirely originates from the same volume of
the crystal and so diffraction intensities corresponding to later times of the pulse may
be perturbed by the earlier X-ray exposure. This is not the case when the crystal is out
of focus, where the mapping of Bragg streaks to time is also a mapping to the position
across the crystal face, and the probe pulse only exposes each diffracting volume for a
time as short as the initial unfocused X-ray pulse. In this case, the velocity-matching
conditions described in Sec.~\ref{sec:matching} still apply, and so if $n_p > 1$, the pump
pulse and the crystal must be tilted appropriately so that the delay is
maintained as a particular X-ray propagates through the thickness of the crystal. This inclination then
gives a variation of the time delay across the face of the crystal and which can be
determined after mapping the Bragg streaks to the incident $\kin$ wave-vectors.

If $n_p = 1$, then velocity matching demands that the pump and probe pulses propagate
parallel to each other. In that situation, it is still possible to create a larger range
of delays across the face of the crystal---by tilting the pulse front of the pump beam
rather than tilting the direction of propagation. A linear pulse front tilt can be
achieved by diffracting the pulse from a plane grating, for example.

A scheme of pump-probe topography using a visible pump and X-ray probe of different
incident angles was first proposed for time-resolved diffraction measurements by Neutze
and Hajdu~\cite{Neutze:1997}, which they called crossed-beam topography. They proposed to
use a very broad collimated X-ray beam and a large crystal, such that the spatial profile
of the crystal is mapped directly onto the detector (without magnification) at each Bragg
peak. For a typical detector pixel size of \SI{75}{\micro\meter}, millimetre-sized
crystals are therefore required to achieve a reasonable number of time bins in a single
measurement. A non-negligible bandwidth is required to ensure that the measured intensity
is fully integrated and to avoid the high sensitivity to lattice distortions which may
cause significant intensity variations across the topogram. Our proposed use of a highly convergent
beam, however, produces a greatly magnified topogram of the crystal that may be hundreds
of pixels wide at the plane of the detector, even for micrometer-sized crystals, allowing
hundreds of time bins in the dataset.

\begin{figure}[tb!]
  \centering
  \includegraphics[width=8cm]{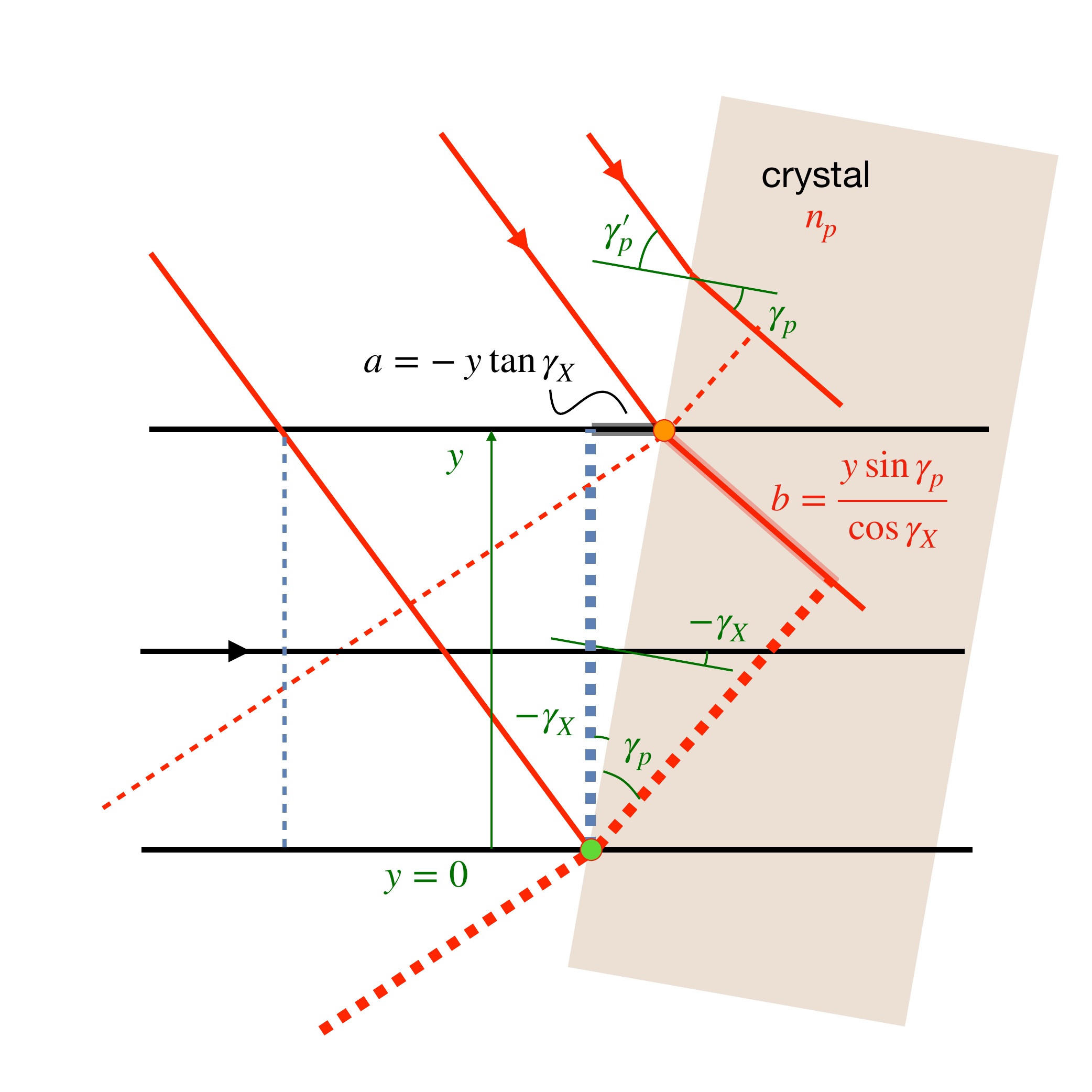}
  \caption{Geometry of crossed-beam topography for the same velocity matching conditions
    as Fig.~\ref{fig:crystal} (b), ignoring the convergence of the X-ray beam. The pump
    pulse (red rays) and X-ray pulse (black rays) coincide in time at the surface of the
    crystal at the green circle at a height $y = 0$. The X-ray pulse arrives later at the
    crystal at height $y$ after a time $T_X(y) =a/c$, at which point the pump pulse has
    already travelled a distance $b$ in the medium of refractive index $n_p$, giving an
    arrival time of $T_p(y) = -b n_p /c$. }
  \label{fig:cross-beam}
\end{figure}

For a thin crystal, the topogram
gives a map of the diffraction efficiency of the crystal projected along the incident
X-ray beam direction~\cite{Lang:1959}. When the crystal face is tilted by $\gamma_X$
relative to the X-ray beam, Eqn.~\ref{eq:delta-t-defocus} is modified to
\begin{eqnarray}
  \begin{aligned}
  \label{eq:6}
  T_X (\phi_x,\phi_y) = \frac{z}{c} \biggl \{&-\phi_y \tan \gamma_X +  \nonumber \\*
  &\left. \phi_y^2 \tan^2 \gamma_X + \frac{\phi^2}{2} \left
      (1-\frac{V\, f}{z} \right) \right\}.
  \end{aligned}
\end{eqnarray}
where $\phi^2 = \phi_x^2+\phi_y^2$. In the general scheme of Fig.~\ref{fig:crystal} (b), the arrival time of the pump on the
surface intersecting with the $\kin$ vector described by $(\phi_x,\phi_y)$ is  
\begin{equation}
  \label{eq:7}
  T_p(\phi_x,\phi_y) = -\frac{z}{c} \frac{n_p \sin \gamma_p}{\cos \gamma_X} \left( \phi_y - \phi_y^2 \tan
    \gamma_X \right). 
\end{equation}
The geometry showing the linear sweeps of the two pulses across the face of crystal as a
function of the transverse position $y = \phi_y \, z$ is given in
Fig.~\ref{fig:cross-beam}.  The pump-probe delay across the face of the crystal relative
to the delay at the optical axis, without any additional pulse-front tilt, can then be
written as
\begin{eqnarray}
  \begin{aligned}
  \label{eq:delta-t-matching}
  T(\phi_x,\phi_y) = \frac{z}{c}\biggl\{& -\phi_y\, G(\gamma_X,\gamma_p) (1- \phi_y \tan
  \gamma_X) +  \nonumber \\
    &\left. \frac{\phi^2}{2} \left (1 -\frac{ V f}{z} \right) \right\} 
    \end{aligned}
\end{eqnarray}
where
\begin{equation}
  \label{eq:9}
  G(\gamma_X,\gamma_p) = \tan \gamma_X -\frac{n_p \sin \gamma_p}{\cos\gamma_X}.
\end{equation}
This term sets the magnitude of the linear component of the map of delays (here in the
$\phi_y$ direction), proportional to the optical path differences highlighted in
Fig.~\ref{fig:cross-beam}. It is seen that the linear component of $T$ is equal to the
time of flight across the transverse extent of the crystal ($z\, \phi_y/c$) multiplied by
the factor $G$, and as such may be the largest contributor to the range of time delays.
It is found that for all incident pump angles $\gamma_p'$, when the tilt $\gamma_X$ is set
to obey the velocity-matching condition of Eqn.~\ref{eq:matching}, Eqn.~\ref{eq:9} reduces
to a form similar to Eqn.~\ref{eq:5} such that
\begin{equation}
  \label{eq:8}
  G(\gamma_X) = \sqrt{n_p^2-1}.
\end{equation}
Therefore, this linear component of $T$ does not depend on the tilt of the crystal nor the
corresponding inclination of the pump, as long as they satisfy the condition of
Eqn.~\ref{eq:matching}. If an off-axis lens is used, such as depicted in
Fig.~\ref{fig:chirp}, this linear term may add to or compensate the quadratic term caused by the
pulse front curvature and defocus.  Also, when $n_p = 1$ then $G = 0$, showing that the
tilt of the crystal does not affect the time delay in that case. Indeed, the shape of the
crystal does not matter when the pump and probe both propagate through the crystal with
the same speed, but the propagation direction of the pump must be set to be colinear
with the X-ray probe. A linear variation in delay across the crystal face can instead be accomplished with a pulse
front tilt of the pump or probe.

In the orthogonal transverse direction to the pump-pulse inclination, the variation in $T$
is quadratic and spans a smaller range of times. Thus the two-dimensional map allows both
a large range in time delays as well as a dense sampling of these times. As an example, a
map of delay times obtained with $\gamma_X =-\SI{41.2}{\degree}$ (as illustrated in
Fig.~\ref{fig:crystal} (a) for $n_p = 1.33$), $f = \SI{1.25}{\milli\meter}$, and
$z = \SI{0.3}{\milli\meter}$ is shown in Fig.~\ref{fig:chirp} (d). Here the delay spans
\SI{50}{\femto\second}.

As with the case of the crystal in focus, the range of tilts $\phi$ of the diverging X-ray beam will
lead to errors in the velocity matching. A change in the angle $\phi$ of the ray relative
to the optical axis can be thought of as a change of the inclination of the crystal
$\gamma_X$. It can therefore be 
compensated by adjusting the tilt of the pump pulse to maintain the velocity-matching
condition of Eqn.~\ref{eq:matching} by making the pump pulse slightly
converging. This requires the defocus distance of the crystal to be much larger than
the crystal thickness so that pump rays illuminating a particular X-ray path do not appreciably
vary in angle.

Yet other pumping schemes can be made that use a linear pulse front tilt. For
example, a constant delay across the entire overlap of the pump and probe beams throughout
the crystal volume could be achieved when $n_p > 1$ by adding a
pulse front tilt to the X-ray beam to match the tilt angle of the crystal and
illuminating it with a pump normal to the surface as in Fig.~\ref{fig:crystal} (a). This
could be achieved with an asymmetric Bragg reflection from a flat crystal monochromator
placed upstream of the lens. Both
pulses would arrive simultaneously at the crystal face and then propagate with the same
velocity component in the direction of the X-ray beam.

\section{Experimental Test---Vitamin B$_\textbf{12}$}
\label{sec:B12}

\begin{figure*}[tb!]
  \centering
   \includegraphics[width=0.8\linewidth]{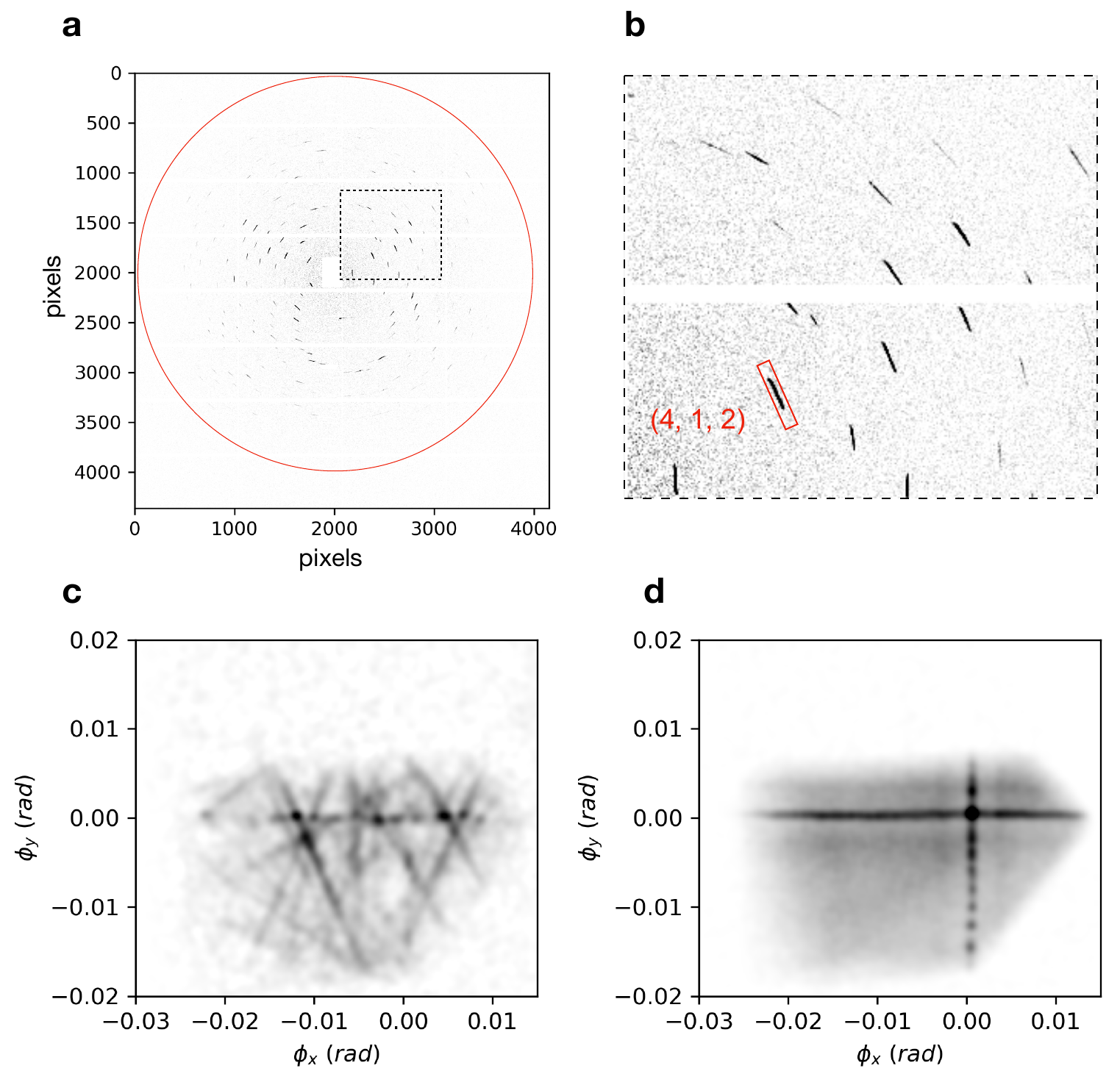}
   \caption[Diffraction from a vitamin B$_{12}$ crystal]{(a) Diffraction from a single
     B$_{12}$ crystal placed \SI{3.5}{\milli\meter} downstream of the focus of a 0.028 NA
     lens at a photon energy of \SI{17.5}{\kilo\eV}. The red circle
     indicates a resolution of \SI{1}{\angstrom}. The region outlined with dashed lines is shown enlarged
     in (b).  (c) Map of the deficit lines obtained from the Bragg streaks of (a). (d) Map
     of the projected crystal diffraction efficiency obtained from integrating the (412)
     Bragg streak, highlighted in (b), over a \SI{2.75}{\degree} fine rotation scan of the
     crystal about the vertical axis.  }
  \label{fig:B12}
\end{figure*}

A convergent-beam diffraction experiment was made using synchrotron radiation to
demonstrate the mapping of Bragg streak intensities to a map of incident $\kin$
vectors. Figures~\ref{fig:B12} (a) and (b) show a convergent-beam diffraction pattern
measured from a vitamin B$_{12}$ (cobalamin) crystal at a photon energy of
\SI{17.5}{\kilo\eV} (wavelength of \SI{0.7}{\angstrom}). Measurements were conducted at
the P11 beamline at the PETRA III synchrotron radiation facility using an X-ray microscope
setup as previously described~\cite{Zhang:2024} and an EIGER X 16M detector with a Si
sensor (Dectris) consisting of $4150 \times 4371$ pixels. The detector was placed
\SI{18.3}{\centi\meter} from the sample, centered on the optical axis, to record a maximum
scattering angle of \SI{53}{\degree} in the detector corner, corresponding to a maximum
resolution of $1/(\SI{0.78}{\angstrom})$.  A pair of on-axis MLLs of 0.028 NA were
oriented orthogonal to each other to focus the beam in the horizontal and vertical
directions with focal lengths of \SI{1.25}{\milli\meter} and \SI{1.26}{\milli\meter},
respectively. The lenses were prepared by masked deposition~\cite{Prasciolu:2015} and each
consisted of \num{16965} bi-layers with a minimum period of \SI{2.06}{\nano\meter}
(manuscript in preparation).  As thick diffractive elements, each lens splits the beam
into a focused order as well as an unfocused order~\cite{Bajt:2018}. While a central
obstruction of the lens paired with a pinhole just upstream of the focus can block all but
the beam focused by both lenses, we instead used an unobstructed lens to ensure that full
$\kin$ maps could be constructed from the Bragg streaks.  A square pinhole of
\SI{10}{\micro\meter} width was placed near the focus to block most of the flux in the
zero order beam combinations, which at the focus form horizontal and vertical line foci
and a collimated beam. The parts of these beams that passed through the pinhole illuminate
the crystal placed downstream of focus and give rise to diffraction additional to the
desired convergent-beam diffraction. Far from this being a disadvantage, we found that the
additional diffraction provides fiducials that assist in determining the coordinates of
the incident deficit lines. The zero order component did require that a beamstop be placed
on axis in front of the detector, however.

The B$_{12}$ crystal was about \SI{140}{\micro\meter} in width and was mounted on a kapton
mesh positioned at $z=\SI{3.5}{\milli\meter}$
downstream of focus. The crystal form has orthorhombic P$2_12_12_1$ symmetry with unit cell parameters
$a=\SI{15.70}{\angstrom}$, $b=\SI{22.15}{\angstrom}$, and $c=\SI{24.96}{\angstrom}$. The
diffraction pattern in Fig.~\ref{fig:B12} was recorded for a single static orientation of
the crystal with an exposure consisting of \num{2.5e9} total incident photons. It is hence
equivalent to one that may be obtained with an attenuated single pulse of an XFEL.  This
pattern was indexed using the known unit-cell parameters, using a procedure related to
that of the pinkIndexer~\cite{Gevorkov:2020} as will be described elsewhere. The indexing
solution provides the rotation matrix of the reciprocal lattice from which the reciprocal
lattice vectors $\vec{q}_{hkl}$ associated with each Bragg streak $\kout(\chi)$ can be
determined. From each of these, the corresponding $\kin(\chi)$ deficit line can thus be
obtained. The deficit lines cut across the lens pupil and hence also across the face of
the crystal.
A map of these deficit lines is given in Fig.~\ref{fig:B12} (c), as a function of the
angular components $\phi_x$ and $\phi_y$ of the $\kin$ vector (see Eqn.~\ref{eq:4})
and uncorrected for the lens pupil
amplitude or structure factors of the reflections. It can be seen that this map is not
complete---there are areas of the crystal that do not lead to strong diffraction---but it
is clear that each point along every measured Bragg streak can be mapped to the particular
$(\phi_x, \phi_y)$ position in the pupil and hence to a particular X-ray arrival time 
$T_X$ (relative to the on-axis beam) as well as to the pump-probe delay time $T$ if
the pump pulse front is well characterised.

To evaluate the accuracy of the mapping of the Bragg streaks to their $\kin(\chi)$ deficit
lines, we also collected a series of diffraction patterns where the crystal was rotated
about the vertical axis over a range of angles comparable to the lens
convergence $2\alpha$. In this case, most Bragg peaks sweep across the entire face of the
crystal to form magnified topograms. We chose the intense $(421)$ reflection, highlighted
in the expanded pattern in Fig.~\ref{fig:B12} (b), to generate such a topogram.  The
diffraction intensity of this streak, integrated over a fine rotation scan of
\SI{2.75}{\degree} of the crystal is shown in Fig.~\ref{fig:B12} (d), after isolating the
streak from others in each frame of the rotation scan. (Accumulating directly on the
detector leads to overlapping topograms from neighbouring reflections.) In the figure, the
crystal shape is more apparent than for that obtained from the still diffraction pattern,
with a variation in diffraction efficiency corresponding predominantly to the projected
thickness of the crystal.

A particularly striking feature of the intensity map of Fig.~\ref{fig:B12} (d) is the
cross pattern, formed from diffraction of the horizontal and vertical line foci and the
collimated beam transmitted by the pinhole. These beams map to $\phi_x = 0$ or
$\phi_y = 0$ and hence provide the absolute coordinate of the $\kin$ map from which the
times $T_X$ can be determined to high accuracy. They provide fiducials that can be observed in single
Bragg streaks and help to confirm the precision of the $\kin$ map of Fig.~\ref{fig:B12}
(c), where the horizontal line is particularly apparent. When two strong peaks of
diffraction intensity are observed along a single Bragg streak, both $\phi_x$ and $\phi_y$
can be directly obtained. We find that the accuracy of the mapping of the deficit lines in
Fig.~\ref{fig:B12} (c) to be better than a single pixel (which has an angular width of
\SI{0.4}{\milli\radian} in this case). The crystal was well over-filled by the diverging
beam which spans an angular range in both directions of $2\alpha$ from \SI{-0.028}{radian}
to \SI{0.028}{radian}. As such, even longer Bragg streaks could have been measured with
the crystal placed closer to focus.

We note that the diffraction intensity originating from  $\phi_x = 0$ or
$\phi_y=0$ does not necessarily map back to
the corresponding coordinate on the face of the crystal. For example, the collimated beam, which diffracts to the
center of the cross, originates from the entire area of the crystal it illuminates
(dependent on the size of the pinhole). In a short-pulse measurement, the intensities in the cross therefore cannot be
used to obtain precise time-resolved data when used with a tilted pump pulse, but they do act as an in-built fiducial to
define the $\phi_x$ and $\phi_y$ coordinates for each streak.  

\section{Temporal Resolution}
\label{sec:resolution}
In Fig.~\ref{fig:B12} (c), measured diffracted intensities of various reflections are
mapped back to the corresponding incident $\kin$ wave-vectors in terms of the angular
components $(\phi_x, \phi_y)$. This shows that together with the spatial distribution of the pump pulse
arrival times, $T_p$, the time delay $T=T_X-T_p$ between pump and probe can be mapped to high accuracy.
To account for jitter between the pump and the probe it may be possible to identify a
particular event time (such as the zero time delay) in each snapshot pattern, requiring
that the measured range exceeds that jitter. The span of $T_X$ is proportional to the
combined dispersion of the lens and crystal defocus distance, and to the square of the
lens NA, whereas the span of $T_p$ depends on the tilt of the pump and the extent of the
crystal in the defocused beam.  Assuming velocity matching of the pump and probe pulses
throughout the thickness of the crystal, the temporal resolution is dictated by how finely
the map of $T$ is sampled, which is given by the angular extent of the detector
pixels. Since we aim for atomic resolution, the detector must be placed close enough to
capture high scattering angles, which then places a lower limit on the pixel angular
extent. Temporal resolution is therefore best maximised by using a detector with a large
number of pixels and with the largest possible lens NA so that the Bragg streaks span as
many pixel time bins as possible. In our B$_{12}$ measurement of Sec.~\ref{sec:B12}, the
beam diverging from the focus, $A(\kin)$, covered $140 \times 140$ pixels. If this
experiment was implemented with XFEL pulses, the arrival times of rays in the plane of the
crystal located \SI{3.5}{\milli\meter} downstream of focus would span
\SI{12.4}{\femto\second}. Due to the quadratic dependence of $T_X$ on $\phi$, the temporal
sampling is not uniform and is finer for earlier times. Although neighbouring pixels may
sample different times by less than \SI{1}{\atto\second}, the temporal resolution is
determined by the range of arrival times of rays incident on the sensitive area of each
pixel. These times vary greatest in the radial direction. In our example, neighbouring
pixels in the radial direction differ by \SI{3}{\atto\second} at the early times for
incident rays near to the optical axis (e.g.\ dark red portions of Bragg streaks of
Fig.~\ref{fig:pattern} (a)) to \SI{0.36}{\femto\second} for the marginal incident rays
(dark blue in Fig.~\ref{fig:pattern} (a)) and the mean resolution is
\SI{0.18}{\femto\second}, as listed in column A of Table~\ref{tab:resolution}. This
corresponds to the time delay map for an on-axis collimated pump pulse with $n_p = 1$ (e.g. a
soft X-ray pulse).
The table
assumes the same detector size of 16 Mpixels as used in our measurement and which is
available at the Bernina beamline of the SwissFEL~\cite{Ingold:2019}.

\begin{table*}[tb]
  \caption[Temporal resolution]{Temporal resolutions for different configurations, all
    with $f=\SI{1.25}{\milli\meter}$ and $\mathrm{NA}=0.028$, at a wavelength of
    \SI{0.7}{\angstrom} and with a centred $4000 \times 4000$ pixel detector positioned to record to a resolution
    of \SI{0.78}{\angstrom} at its corner and a pixel angular separation of
    \SI{0.8}{\milli\radian}. The time resolution $\Delta T$ is calculated as the variation in delays across
    the active area of a detector pixel. }
  \label{tab:resolution}
  \begin{ruledtabular}
    \begin{tabular}{cSSSS}
    & \textbf{A} & \textbf{B} & \textbf{C} & \textbf{D} \\
    \cline{2-5}
    & {Experiment} & {Near focus} &  {Off-axis} & {Cross beam} \\
    \hline
     $z$ (mm) & 3.5 & 0.18 & 0.18 & 0.18 \\
    Beam width ($\mu$m) & 200 & 10 & 10 & 10 \\
    Min. $\phi$ (rad)  & 0     & 0         & 0.014 & 0 \\
    Max. $\phi$ (rad) & 0.04 & 0.04 & 0.091 & 0.04 \\
    $n_p$ & 1.0 & 1.0 & 1.0 & 1.33 \\
    $T$ (fs) & 12.4& 3.7 & 19.2 & 31.0\\
    Min.\ $\Delta T$ (fs)          & 0.003 & 0.001 & 0.04 & 0.16\\
    Max.\ $\Delta T$ (fs)         & 0.36 & 0.11   & 0.25   & 0.32 \\
    Mean $\Delta T$ (fs)          & 0.18 & 0.05   & 0.14   & 0.24 \\
      \end{tabular}
  \end{ruledtabular}
\end{table*}

The \SI{140}{\micro\meter} B$_{12}$ crystal used above is larger than required for serial
crystallography.
Column B in Table~\ref{tab:resolution} corresponds to a defocus of \SI{180}{\micro\meter}
where the beam width is \SI{10}{\micro\meter}, suitable for micro-crystals and giving a
\SI{3.7}{\femto\second} span of X-ray probe arrival times.
For a given lens focal length, a larger range of X-ray arrival times can be achieved with an off-axis lens, as listed in the
table as column C. Here, the bi-layer periods in the lenses were taken to decrease from
\SI{7.0}{\nano\meter} to \SI{1.1}{\nano\meter} such that the deflection angles from each
range from \SI{0.010}{\radian} to \SI{0.066}{\radian}. Along the diagonal of the pupil of
the composite lens the deflection angle $\phi$ ranges from \SI{0.014}{\radian} to
\SI{0.091}{\radian}, giving a time span of \SI{19.2}{\femto\second} at the near-focus position.

Crossed-beam topography gives much greater control on both the range and resolution of
times. For example, by inclining a collimated pump pulse to the optical
axis of the X-ray lens as depicted in Fig.~\ref{fig:chirp} (c), or by tilting the plane of
a wide thin crystal, the range of delays can
either be reduced or increased, depending on the direction of the inclination. Column D in
Table~\ref{tab:resolution} lists the timing characteristics for the on-axis lens
configuration but with the pump pulse inclined to give a linear sweep of
\SI{30}{\femto\second}. This produces a near-linear variation of the delay time across the
$\kin$ map. 

Other possibilities exist: for example, a quadratic dependence of arrival times of the
pump beam could be achieved by focusing the pump beam with a dispersive lens, which could
further reduce the range of delays across the beam and compensate for the variation of
angles $\gamma_X$ due to the X-ray beam curvature.
In practise the achievable resolution will be limited by how accurately Bragg streaks can
be mapped back to the $\kin$ space of the pupil, which may depend on lens aberrations and
changes in beam pointing, as well as signal to noise of the measurements. Our
experimental tests indicate that the error is certainly less than 2 pixels, corresponding to the pixel
sampling assumed in Table~\ref{tab:resolution}. As mentioned in the Introduction, the
temporal resolution also depends on the duration of the pump pulse, which must necessarily
be of short wavelength to reach sub-femtosecond duration. EUV or soft X-ray pulses might
be generated by an independent laser-driven source or emitted by the FEL in a two-colour
scheme. In such cases the electron dynamics would be initiated by direct photoionisation
and so may differ from dynamics of chromophores and electron transfer. This may open
opportunities to target specific atomic species as well as to directly study such
processes as inter-atomic Coulombic decay~\cite{Santra:2001} and ultrafast charge
transfer~\cite{Rudenko:2017}. It may also be possible that a phase-stable single-cycle optical pulse
triggers electron dynamics at a specific point in that cycle, akin to the generation of
attosecond pulses in the HHG process.

\section{Conclusions}
\label{sec:conclusion}
Convergent-beam serial crystallography, utilising high-NA lenses such as MLLs, offers the
control and precision for acquiring high-resolution diffraction data with sub-femtosecond
temporal resolution.
A potential drawback of convergent-beam diffraction, shared with Laue diffraction, is that
the signal to background ratios of reflections are lower than for a collimated beam, since
diffraction intensity is spread over many more pixels. This is a consequence of the much
greater information content of the snapshot diffraction pattern. Nevertheless, as described in this
paper, the use of high-NA X-ray lenses offers several compelling advantages for
time-resolved serial crystallography:
\begin{itemize}
\item Diffraction patterns consist of Bragg streaks, along which time can be directly
  encoded, and decoded by indexing the pattern to determine the span of incident wave
  vectors $\kin(\chi)$ associated with each streak.
\item In addition to the two-dimensional focus, paired MLLs give rise to two orthogonal line foci and a collimated
  beam which all can be selected with an aperture to illuminate the crystal and provide
  in-built fiducials that provide the absolute angular coordinate (and hence arrival time)
  of the incident rays.
\item The X-ray probe is focused to a small probe size, bringing all available flux of an
  XFEL pulse to interact with micrometer-sized crystals.
\item The participating reciprocal-space volume is greatly increased by a high-NA
  convergent beam, compared with a collimated beam, giving rise to a large number of Bragg
  streaks in a snapshot diffraction pattern. For crystals of small unit cell dimensions
  this avoids the sparsity of collimated-beam diffraction patterns that make them
  difficult to index, and greatly reduces the number of snapshot patterns required to
  acquire a complete dataset. The participating volume is much larger than attainable by
  Laue diffraction at XFEL sources.
\item The Bragg streaks provide predominantly fully integrated intensities, further
  reducing the number of patterns required to obtain precise estimates of structure
  factors.
\item When the crystal is placed in focus, all diffraction originates from a sub-volume of
  the crystal which may have a transverse extent of only a single unit cell. When out of
  focus, the Bragg streaks map across the face of the crystal and can be used to obtain a
  projection image of the crystal's diffraction efficiency. This map can be used to
  account for the effects of any variations in crystal shape or efficiency on the
  Bragg-streak intensities.
\item The obtained projection image of the crystal is highly magnified, enabling
  crossed-beam topography of small crystals with high temporal resolution.
\item In crossed-beam topography, the pump and probe pulses, propagating with differing speeds due to unequal
  refractive indices, can maintain a constant delay over the path of the X-ray
  beam through the crystal thickness by appropriately tilting the pump beam and
  crystal face, as prescribed by the velocity-matching condition.
\item As long as the velocity-matching condition is satisfied, the span of time delays in
  crossed-beam topography is independent of the crystal tilt and is set by the transverse
  extent of the defocused X-ray beam size at the crystal and the refractive index of the
  crystal medium.
\item The wavefront control achieved by focusing a pulse to a nanometer spot is synonymous
  with small path-length errors that are required for sub-femtosecond timing.
\end{itemize}

\begin{acknowledgments}
We thank Francesca Calegari, Robin Santra, TJ Lane, Alke Meents and Miriam Bartelmess for
discussions. We acknowledge DESY (Hamburg, Germany), a member of the Helmholtz Association
HGF, for the provision of experimental facilities and funding. Parts of this research were
carried out at PETRA III and we would like to thank Johanna Hakanp\"{a}\"{a} and her
colleagues for assistance in using the P11 beamline. Beamtime was allocated for proposal
I-20231166. This work was further supported by the Cluster of Excellence ``CUI: Advanced
Imaging of Matter'' of the Deutsche Forschungsgemeinschaft (DFG) - EXC 2056 - project ID
390715994.
\end{acknowledgments}

\section*{Data Availability Statement}
Raw data were generated at the PETRA III synchrotron radiation facility. Derived data
supporting the findings of this study are available from the corresponding author upon
reasonable request.

\bibliography{CB-atto}

\end{document}